\documentclass[%
 reprint,
superscriptaddress,
 amsmath,amssymb,
 aps,
]{revtex4-2}

\usepackage{graphicx}
\usepackage{dcolumn}
\usepackage{bm}
\usepackage{mathtools} 
\usepackage{microtype}
\usepackage{siunitx}
\usepackage{color}
\renewcommand{\figurename}{Figure}

\definecolor{webgreen}{rgb}{0,.5,0}
\usepackage[colorlinks=true,citecolor=webgreen]{hyperref}

\definecolor{violet}{rgb}{0.4,0.2,0.8}

\definecolor{orange}{rgb}{0.9,0.5,0.2}

\begin{document}
\preprint{APS/123-QED}

\title{A model for the fragmentation kinetics of crumpled thin sheets}

\author{Jovana Andrejevic}
\affiliation{John A.~Paulson School of Engineering and Applied Sciences, Harvard University, Cambridge, MA 02138, USA}
\author{Lisa M. Lee}
\affiliation{John A.~Paulson School of Engineering and Applied Sciences, Harvard University, Cambridge, MA 02138, USA}
\author{Shmuel M. Rubinstein}
\affiliation{The Racah Institute of Physics, The Hebrew University of Jerusalem, Jerusalem 91904, Israel}
\author{Chris H. Rycroft}
\email[]{To whom correspondence should be addressed. E-mail: chr@seas.harvard.edu}
\affiliation{John A.~Paulson School of Engineering and Applied Sciences, Harvard University, Cambridge, MA 02138, USA}
\affiliation{Computational Research Division, Lawrence Berkeley Laboratory, Berkeley, CA 94720, USA}


\begin{abstract}
\noindent\textbf{Abstract:} As a confined thin sheet crumples, it spontaneously segments into flat facets delimited by a network of ridges. Despite the apparent disorder of this process, statistical properties of crumpled sheets exhibit striking reproducibility. Experiments have shown that the total crease length accrues logarithmically when repeatedly compacting and unfolding a sheet of paper. Here, we offer insight to this unexpected result by exploring the correspondence between crumpling and fragmentation processes. We identify a physical model for the evolution of facet area and ridge length distributions of crumpled sheets, and propose a mechanism for re-fragmentation driven by geometric frustration. This mechanism establishes a feedback loop in which the facet size distribution informs the subsequent rate of fragmentation under repeated confinement, thereby producing a new size distribution. We then demonstrate the capacity of this model to reproduce the characteristic logarithmic scaling of total crease length, thereby supplying a missing physical basis for the observed phenomenon.
\end{abstract}

\maketitle

\section*{\label{sec:Introduction}Introduction\protect}
\begin{figure}[ht!]
\centering
\includegraphics[width=8.65cm]{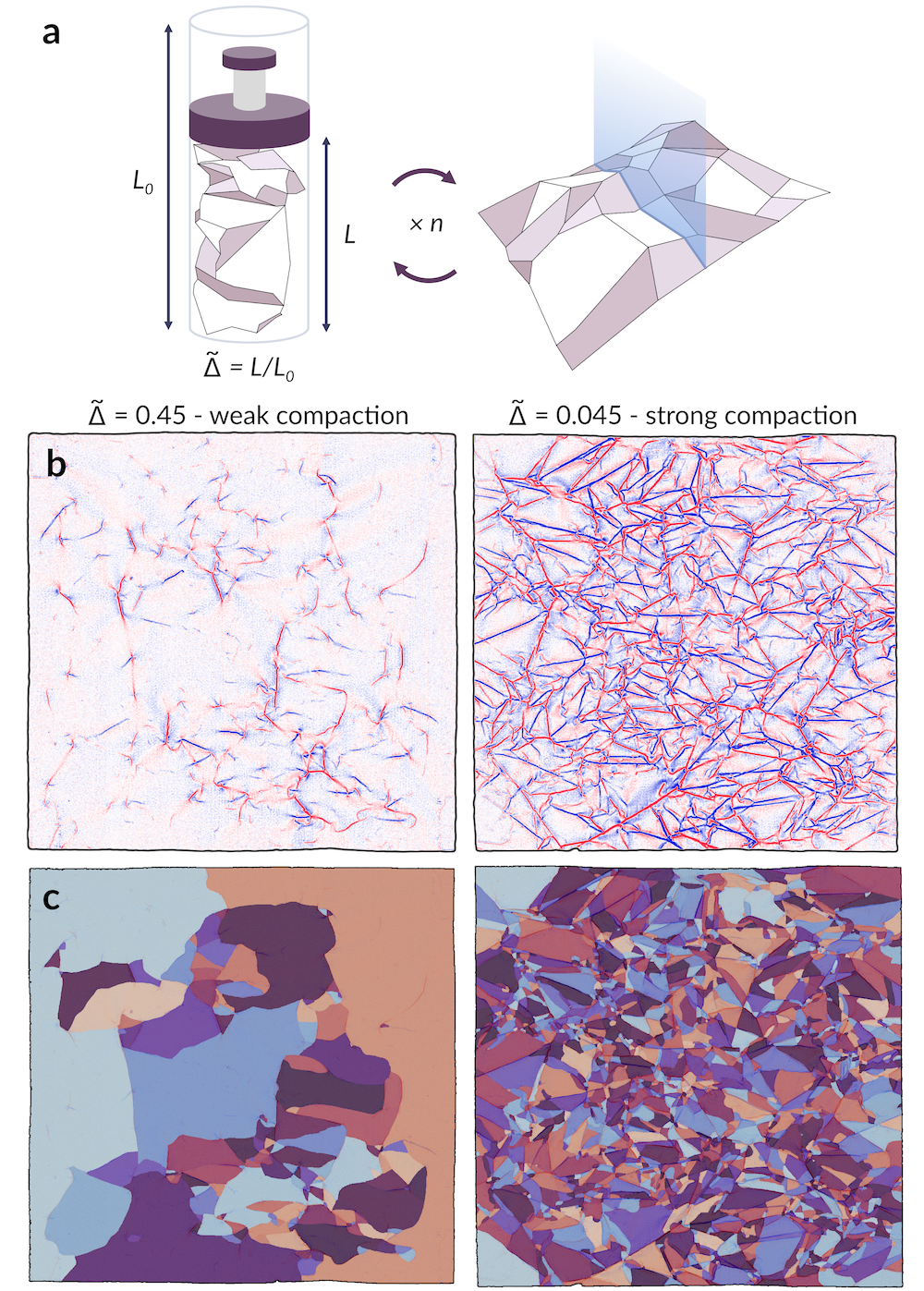}
\caption{\textbf{Data processing.} \textbf{a} An $L_0\times L_0$ Mylar sheet is uniaxially compressed to a compaction ratio $\tilde\Delta=L/L_0$, unfolded, and its height profile scanned using a laser profilometer, for $n$ iterations. \textbf{b} The mean curvature obtained from the height profiles of two distinct sheets at different $\tilde{\Delta}$. Red and blue colors denote folds in opposite directions. \textbf{c} The facet segmentation of \textbf{b}, colored randomly to visually discern facets.}\vspace{-1em}
\label{sample}
\end{figure}

Crumpling is a complex, non-equilibrium process arising in diverse systems across a wide range of length scales, from the microscopic crumpling of graphene membranes~\cite{zang2013multifunctionality}, to the macroscopic folding of Earth's viscoelastic crust~\cite{beloussov1961origin}. Crumpled structures are highly porous, providing function for applications such as high-performance batteries and supercapacitors by increasing the electrochemical surface area~\cite{song2016advanced, wen2012crumpled}. Controlled crumpling  has also been used to tune electronic, optical, and surface properties in graphene films~\cite{zang2013multifunctionality}. Further, understanding the mechanics of crumpling is essential as flexibility and shape conformation become integral considerations in the design of thin, wearable devices~\cite{kaltenbrunner2013ultra, white2013ultrathin}. Despite its ubiquity, a complete understanding of crumpling dynamics remains elusive due to the complexity of the disordered process. Nevertheless, statistical properties of crumpled geometries are highly reproducible in experiment~\cite{deboeuf2013compaction, deboeuf2013comparative, andresen2007ridge, blair2005geometry, balankin2006intrinsically, balankin2008entropic, balankin2006intrinsically, balankin2013fractal} and confirmed via simulation~\cite{vliegenthart2006forced, sultan2006statistics}, which suggests that this complex process is strongly dictated by universal aspects of thin sheets such as topology and self-avoidance~\cite{balankin2013fractal}.

Similarly adopting a coarse-grained perspective, Gottesman \textit{et al.}~\cite{gottesman2018state}  revealed an unexpected order to ridge network evolution in crumpled sheets. By performing a protocol of repeated compaction and unfolding, as in the schematic of Fig.~\ref{sample}a, they demonstrated that the intricate details of ridge networks in crumpled sheets could be reduced to a single collective quantity, the total crease length, which evolves robustly as a logarithm in the number of crumpling repetitions across varying degrees of compaction. Notably, the incremental damage added upon re-crumpling the sheet was found to be independent of the sheets' crumpling history -the sequence of prior compactions performed to produce the current crease network. Rather, the added crease length is determined only by the current total crease length and the new compaction depth. While processes that evolve logarithmically in time are readily observed in a variety of disordered physical systems, including stress or strain relaxation of a compacted sheet~\cite{lahini2017nonmonotonic, balankin2011slow, matan2002crumpling}, conductance relaxation of disordered electronic systems~\cite{eisenbach2016glassy}, and creep dynamics of granular suspensions~\cite{berut2017creeping}, the emergence of a logarithmic model in the specific context of damage evolution in crumpled sheets is clearly distinct, and has had limited physical justification thus far.

In this work, we take a novel approach to characterize crumpling and offer explanation for the logarithmic model by drawing a correspondence between crumpling and fragmentation processes. Fragmentation models have a rich history of theoretical development~\cite{rosin1933laws, kolmogoroff1941logarithmisch, grady2017physics} as well as industrial applications~\cite{rosin1933laws, brown1995derivation} and use in modelling collision and fracture phenomena~\cite{brown1989theory}. Here we concentrate on a theoretical, physically-based rate equation for modeling time-dependent fragmentation detailed by Cheng and Redner~\cite{cheng1990kinetics}, which provides a general framework for processes that may be treated as successive, homogeneous breakups instigated by non-local stresses. The model has been flexibly applied to describe polymer degradation~\cite{wang1995continuous} and volcanic fragments expelled in an eruption~\cite{kaminski1998size}, for example, though to the best of our knowledge this is the first application of such concepts to describe crumpling.

Our work is organized as follows: We derive a scaling solution to the rate equation presented in Ref.~\cite{cheng1990kinetics} which decomposes into a time-invariant distribution of scaled facet area and a time-dependent evolution of mean facet size. We demonstrate that the derived area distribution effectively reproduces key statistical features of experimental crumpled patterns. Fragment distributions are a natural point of comparison between theory and experiment; however, in this work we go a step further to draw additional correspondence in the temporal evolution of the patterns. The temporal parameter that chronicles the evolution of mean facet size serves as an intrinsic clock measuring the maturity of the fragmentation process. We connect this to experimental parameters driving fragmentation forward, namely the number of crumpling iterations and compaction strength which characterize the experiments of Ref.~\cite{gottesman2018state}. To do so, we construct a simple geometric model that likens crumpling to a random walk and is informed by the statistical properties of the derived area distribution. We derive an analytical relation for how geometric frustration occurring in a confined random walk instigates new damage and advances the temporal measure of fragmentation maturity. We demonstrate how this approach allows one to recover the logarithmic evolution of damage in ridge networks observed in Ref.~\cite{gottesman2018state} and explain the history independence of damage formation, thereby furnishing a missing physical basis for this unexpected result.

The key idea behind our model is the extension of fragmentation theory to incorporate a feedback loop: As facets become smaller, they make the sheet more compliant and therefore lower the rate of subsequent fragmentation. This idea may extend to many physical systems where the accumulation of damage inhibits further damage from occurring. Our work therefore shows how fragmentation theory could be applied more generally, and suggests that the universal damage evolution seen in crumpling may have analogs in other physical systems.

\section*{\label{sec:Results}Results\protect}
The collection and processing of experimental crumpling data used to verify analytical results presented in this work is fully detailed in the Methods section. Crease patterns obtained from uniaxially compressed Mylar sheets as shown in Fig.~\ref{sample}b are carefully segmented into individual facets as in Fig.~\ref{sample}c. The samples collected vary in compaction ratio $\tilde{\Delta}$, the ratio of final to initial height, and in the number of successive crumples of the same sheet, $n$. A total of $24$ segmented crease patterns is analyzed spanning $7$ different compaction ratios and including $n=1,2,3$ and $24$ crumpling iterations.

Throughout this work, we will refer to fragmentation in the context of crumpling as the successive partitioning of a thin sheet into smaller, flat facets separated by ridges or creases. To facilitate the construction of a model for this process, we begin with the general theory of fragmentation kinetics outlined in Ref.~\cite{cheng1990kinetics}. Let $x$ represent facet area and $c(x,t)$ the concentration of facets of area $x$ at time $t$; then the linear integro-differential equation describing the evolution of $c(x,t)$ is given by
\begin{equation}\label{frag_eqn}
\frac{\partial{c}(x,t)}{\partial{t}} = -r(x)c(x,t) + \int_x^{\infty}c(y,t)r(y)f(x|y)dy,
\end{equation}
where the effective time $t$ measures the progress or maturity of the fragmentation process, $r(x)$ is the overall rate at which a facet of area $x$ fragments, and $f(x|y)$ is the conditional probability that $x$ is produced from the breakup of $y$, with $y\geq x$. Inferred from this formulation are the assumptions that fragmentation occurs via a homogeneously applied external force, and independently of a facet's shape.

\subsection*{\label{subsec:Breakup rates}Breakup rates\protect}
\begin{figure*}
\centering
\includegraphics[width=\textwidth]{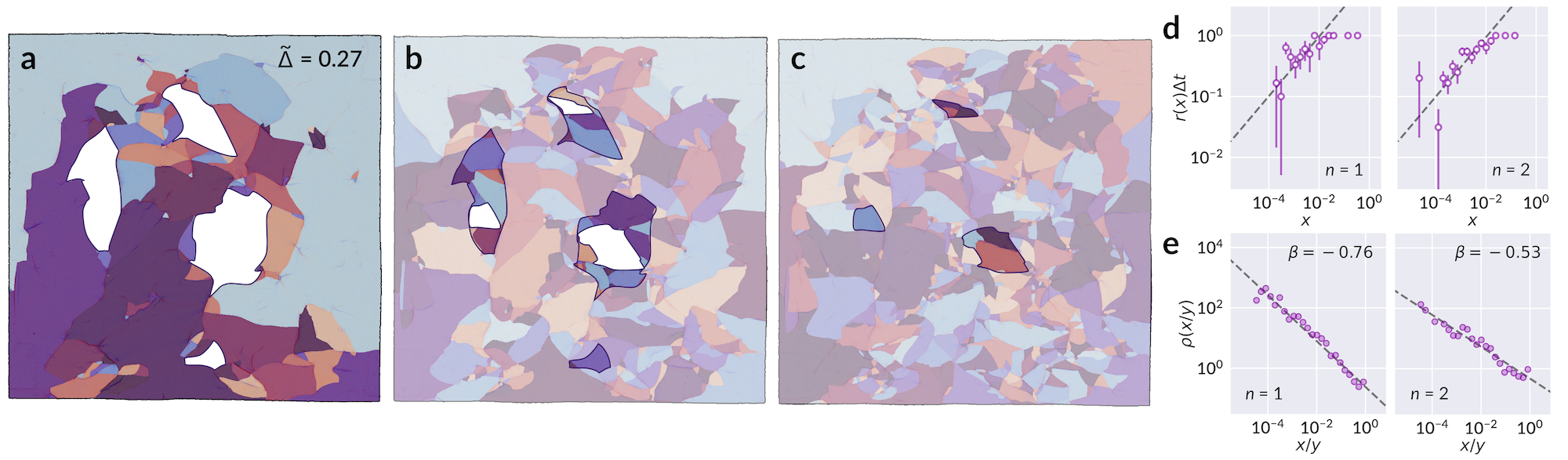}
\strut\vspace{-1.8em}
\caption{\textbf{Estimation of breakup rates.} \textbf{a} Segmentation of a sample sheet crumpled once at compaction $\tilde{\Delta}=0.27$, with four selected facets outlined and emphasized in white. Shown in \textbf{b} are the new facets that subdivide those regions after crumpling a second time ($n=2$). In \textbf{c}, the subdivision of the facets highlighted in \textbf{b} is shown after $n=3$. \textbf{d} For $n=1$ (left), the proportion of facets $r(x)\Delta{t}$ present in \textbf{a} as a function of their area $x$, which have fragmented into at least two distinct facets in \textbf{b} over the elapsed $\Delta{t}$ between $n=1$ and $n=2$. For $n=2$ (right), the fraction of facets from \textbf{b} which have fragmented in \textbf{c}. Error bars denote the standard deviation of the fragmentation probability if the fragmentation of each facet is regarded as a Bernoulli trial, with the fraction of fragmented facets taken as the success probability within each histogram bin. The dashed line corresponds to $\sqrt{x}$. \textbf{e} The probability density function $\rho(x/y)$ of facet areas $x$ normalized by their parent facet's area $y$ from the previous crumpling iteration. The $n=1$ (left) panel is the distribution of area fractions for facets in \textbf{b} relative to their parent facets in \textbf{a}, and $n=2$ the corresponding distribution for facets in \textbf{c} relative to \textbf{b}. The dashed line corresponds to a fit of Equation~\eqref{rho} with the fitted exponent $\beta$ given.}
\label{rates}
\end{figure*}

In order to assess the correspondence between crumpling and a fragmentation process as described by Equation~\eqref{frag_eqn}, two relationships must be specified: the overall breakup rate $r(x)$, and conditional breakup probability $f(x|y)$, which characterize fragmentation at the scale of an individual facet. Two principles help shape our formulations of the two: Firstly, a common choice of $r(x)$ consistent with physical breakup processes is the homogeneous kernel $r(x) = x^{\lambda}$~\cite{cheng1990kinetics}. Furthermore, the conditional probability $f(x|y)$ must satisfy area conservation:
\begin{equation}
    \int_0^y xf(x|y)dx = y. \label{area_cons}
\end{equation}
We use the collection of facets within each sheet as representative samples from which breakup rates may be determined. Fig.~\ref{rates}a-c shows a typical example over three crumpling repetitions and traces the progressive fragmentation of selected facets. From such sequences, we estimate $r(x)$ by determining the fraction of facets which fragment between two successive crumples as a function of their area $x$. The rates are computed separately for each sheet to ensure the same change in $t$ elapses for all facets considered at a time. Without loss of generality, the values of $x$ in all results are scaled so that \SI{10}{\centi\metre} $\times$ \SI{10}{\centi\metre}, the size of one sheet, corresponds to unit area. A breakup rate of the form $r(x) = x^\lambda$ appears consistent with experimental breakup data, as shown in Fig.~\ref{rates}d. Results for samples at other compaction ratios $\tilde{\Delta}$ are provided in Supplementary Fig.~\ref{rx_full}. We note one limitation of this analysis: Sheets crumpled at a low compaction ratio may have too few facets for a robust sample size from which to infer a strong statistical trend; in the opposing extreme, sheets at high compaction likely undergo a cascade of multiple fragmentation events in a single crumpling iteration, and thus obscure the statistics of single breakup events. The power law relationship is motivated both by its consistency predominantly at low compaction, as well as the simplicity it affords later in our model.

To deduce $f(x|y)$ it is helpful to first examine the distribution $\rho(x/y)$ of the area fraction $x/y$ that a child facet occupies relative to its parent. That is, if $x$ is the area of a facet at crumpling iteration $n$, and $y$ the area of its enclosing facet at iteration $n-1$, then $\rho(x/y)d(x/y)$ is the probability that a facet breaks to produce a fragment that is between $x/y$ and $x/y+d(x/y)$ of its initial area, for a small differential element $d(x/y)$. To account for minor misalignment between successive scans, a child facet is identified if at least half of its area lies within the contour of the candidate parent facet. The area fractions display a power law distribution, as shown in Fig.~\ref{rates}e, and suggest a fit to a probability density function of the form
\begin{equation}
\rho\bigg(\frac{x}{y}\bigg) = (\beta+1)\bigg(\frac{x}{y}\bigg)^{\beta}, \label{rho}
\end{equation}
supported on $x/y\in [0,1]$. This formulation introduces the assumption that fragmentation is a scale invariant process; while this is consistent with the present data, we note that a physical lower limit on facet area exists, and would expect deviation from scale invariant behavior as facet areas become comparable to the sheet thickness. Nevertheless, we observe clear indication of a power law relationship within our data, as shown in Fig.~\ref{rates}e. Extended results are provided in Supplementary Fig.~\ref{fxy_full}; as previously noted, samples at high compaction undergo a succession of fragmentation events between crumples, and thus their distributions begin to depart from the power law dependence toward the more mature facet distributions of repeatedly crumpled sheets we present later, in our discussion of ridge length statistics.
Taking $f(x|y)$ proportional to $\rho(x/y)$ and obtaining the appropriate normalization which satisfies Equation~\eqref{area_cons}, we arrive at our final forms for the breakup rates:
\begin{subequations}
\begin{align}
r(x) &= x^{\lambda}, \\
f(x|y) &= \frac{1}{y}\bigg(\frac{\beta+2}{\beta+1}\bigg)\rho\bigg(\frac{x}{y}\bigg) = \frac{1}{y}(\beta+2)\bigg(\frac{x}{y}\bigg)^{\beta}.
\end{align}
\end{subequations}
It will prove useful to express the free parameter $\beta$ as
\begin{equation}
\beta = \frac{a}{2}-1.
\end{equation}
With this definition, we demonstrate in a following subsection that the new free parameter $a$ corresponds to the shape parameter for the distribution of crease length.

\subsection*{\label{sec:Scaling solution}Scaling solution\protect}
With $r(x)$ and $f(x|y)$ specified, we pursue an analytical solution to the fragmentation rate equation, Equation~\eqref{frag_eqn}. Specifically, we seek a scaling solution independent of initial conditions, a property that allows us to solve analytically and proves compatible with the chosen form of homogeneous breakup kernels~\cite{cheng1990kinetics}. We thereby test a scaling ansatz $c(x,t) = \phi(\xi)/s(t)^2$ as proposed in Ref.~\cite{cheng1990kinetics}, where $\xi=x/s(t)$, and the mean area, $s(t)$, carries all explicit dependence on $t$. The scaling function $\phi(\xi)$ satisfies $\int_0^{\infty}\phi(\xi)d\xi = 1$ and $\int_0^{\infty}\xi\phi(\xi)d\xi = 1$ such that $\int_0^{\infty}c(x,t)dx =1/s(t)$ gives the average number of fragments, and $\int_0^{\infty}xc(x,t)dx = 1$ is the total area,
conserved by construction. We note that $\phi(\xi)$ is a valid probability density function and represents the distribution of the scaled facet area $\xi$. The rate equation may be solved following the procedure in Ref.~\cite{cheng1990kinetics} as detailed in Supplementary Note 1; by this approach we arrive at a solution
$c(x,t) = \phi(\xi)/s(t)^2$, valid at large $t$, with
\begin{subequations}
\begin{align}
\phi(\xi) &= \frac{\lambda}{\Gamma\Big(\frac{a}{2\lambda}\Big)}G(a,\lambda)\big(G(a,\lambda)\xi\big)^{\frac{a}{2}-1}e^{-\big(G(a,\lambda)\xi\big)^\lambda}, \label{phi_gen} \\
s(t) &= G(a,\lambda)t^{-1/\lambda}, \label{mean_gen}
\end{align}
\end{subequations}
where $G(a,\lambda) =\Gamma\big(\frac{a+2}{2\lambda}\big)/\Gamma\big(\frac{a}{2\lambda}\big)$, and $\Gamma(z)$ is the gamma function. We motivate a fixed choice of the breakup rate parameter $\lambda=1/2$ both by its consistency with breakup statistics at low compaction, which more accurately reflect single breakup events as discussed earlier, as well as the simplification it provides to obtain an analytically tractable model. We thereby obtain the final forms
\begin{subequations}
\begin{align}
\phi(\xi) &= \frac{a(a+1)}{2\Gamma(a)}\big(a(a+1)\xi\big)^{\frac{a}{2}-1}e^{-\sqrt{a(a+1)\xi}}, \label{phi} \\
s(t) &= \frac{a(a+1)}{t^2}. \label{mean}
\end{align}
\end{subequations}

\subsection*{\label{subsec:Ridge length statistics}Ridge length statistics\protect}
\begin{figure*}
\centering
\includegraphics[width=\textwidth]{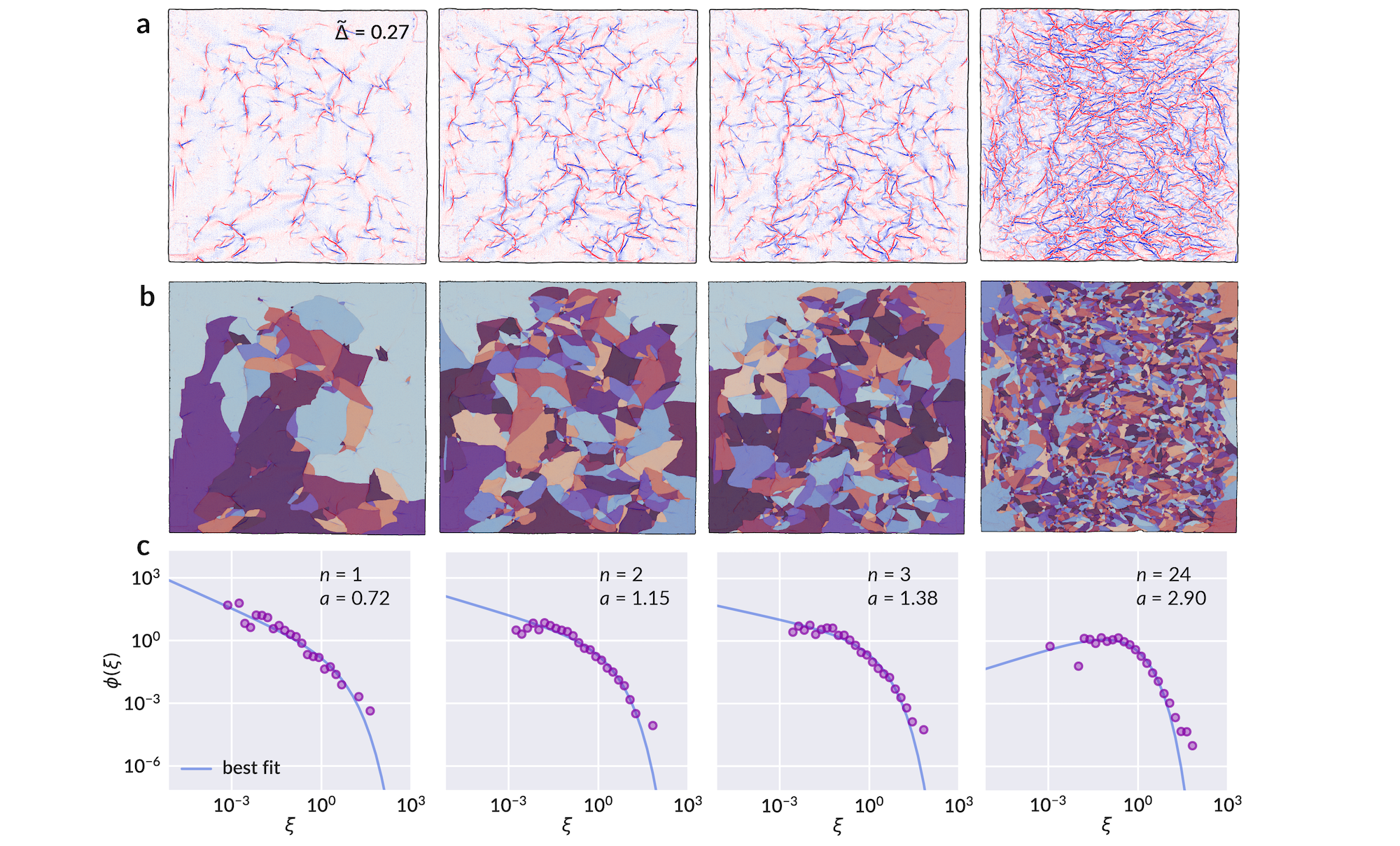}
\caption{\textbf{Facet area distributions for a sample sheet.} \textbf{a} The map of mean curvature for iterations $n=1,2,3$ and $24$ of a sample sheet crumpled with compaction ratio $\tilde{\Delta}=0.27$, and \textbf{b} the corresponding facet segmentation. \textbf{c} Experimental distributions of scaled facet area $\xi=x/s$ for each sample (scattered points) and best fit curve to Equation~\eqref{phi} (solid line). The parameter $a$ for each sample is obtained via self-consistent calculation by Equations~\eqref{mean} \&~\eqref{at_eqn}, and only the universal parameter $\tau\approx{24.041}$ is collectively fit for all samples.}
\label{distribution}
\end{figure*}

\begin{figure}
\centering
\includegraphics[width=8.6cm]{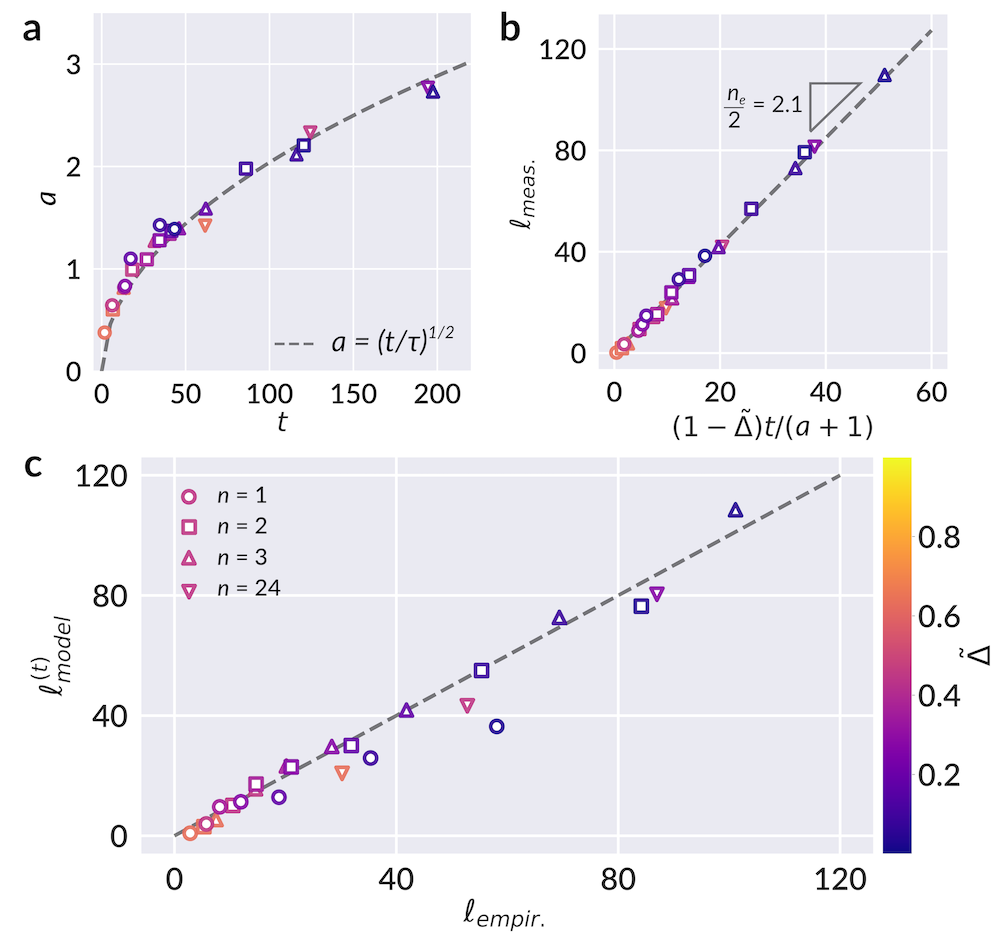}
\caption{\textbf{Model parameters and preliminary comparison to empirical result.} \textbf{a} Individual fits of the shape parameter $a$ from Equation~\eqref{phi} for each facet distribution (scattered points) alongside the best fit to Equation~\eqref{at_eqn} (dashed line), corresponding to $\tau\approx 24.041$. \textbf{b} The measured total crease length $\ell_{\text{meas.}}$ of each segmented sheet plotted against the quantity $(1-\tilde{\Delta})t/(a+1)$ (scattered points). By Equation~\eqref{eqn_of_state_model}, we expect the slope of this plot to correspond to $n_e/2$, or half the average number of facets per facet. A best fit line reveals $n_e/2\approx 2.1$, or $n_e\approx 4.2$ (dashed line). \textbf{c} With the results of \textbf{a} and \textbf{b}, we can make a comparison of $\ell_{\text{model}}^{(t)}\equiv\ell(t,\tilde{\Delta})$ as given by the derived relation Equation~\eqref{eqn_of_state_model}, with the experimental model $\ell_{\text{empir.}}\equiv\ell(n,\tilde{\Delta})$ of Equation~\eqref{eqn_of_state}. The parameters of Equation~\eqref{eqn_of_state} are set to $c_1 = 52$ (normalized by $100$ mm sheet size) and $c_2 = 0.1$,
comparable to the best fit values reported in Ref.~\cite{gottesman2018state}: $c_1 = 5200$ mm,
$c_2 = 0.063$. A $1:1$ reference line (dashed) is provided as a guide to the eye, and shows good agreement between the two models. Marker colors in all panels correspond to different values of $\tilde\Delta$, as indicated by the colorbar.}
\label{log_model_compare}
\end{figure}

To facilitate comparison with $\phi(\xi)$ in Equation~\eqref{phi}, the area of individual facets is scaled by the mean area for that sheet and plotted as a histogram using logarithmically spaced bins. Fig.~\ref{distribution} shows the mean curvature, hand segmentation, and scaled area distributions for a typical example from our dataset at $4$ different crumpling iterations $n$. By our preliminary observations from Fig.~\ref{rates}e, we notice sample-to-sample variation in the parameter $\beta$ (correspondingly $a$), which suggests $a$ is a function of $t$; however we expect weak dependence on $t$ such that $\lim_{t\rightarrow\infty}da/dt=0$, to uphold the assumptions made in solving Equation~\eqref{frag_eqn}. Indeed, an individual fit of $a$ to each distribution of facet areas reveals a dependence of the form
\begin{equation}
    a(t)=\sqrt{t/\tau} \label{at_eqn}
\end{equation} with a universal parameter $\tau$, as shown in Fig.~\ref{log_model_compare}a. Thus, Fig.~\ref{distribution}c additionally shows a best fit curve to Equation~\eqref{phi} with $\tau\approx 24.041$ as a universal fitting parameter across all samples, and individual $a$ and $t$ for each sample computed by solving Equations~\eqref{mean} \&~\eqref{at_eqn} self-consistently. The complete set of segmented crease patterns and fitted area distributions for all data samples is provided in Supplementary Figs.~\ref{data} \&~\ref{fit}.

The close correspondence between Equation~\eqref{phi} and experimental data supports the hypothesis that successive partitioning of the sheet's surface into facets during crumpling evolves according to the fragmentation process described by Equation~\eqref{frag_eqn}. We can study the further implications of this statistical description on attributes such as the distribution of crease length, which has been explored in previous studies~\cite{deboeuf2013compaction, deboeuf2013comparative, andresen2007ridge, blair2005geometry, vliegenthart2006forced, sultan2006statistics, balankin2015edwards}. Let $X$ be the random variable representing the area of a single facet. Following from Equation~\eqref{phi}, $X$ is distributed as
\begin{equation}
f_X(x) = \frac{1}{2\theta^2\Gamma(a)}\bigg(\frac{x}{\theta^2}\bigg)^{\frac{a}{2}-1}e^{-\sqrt{x}/\theta}, \label{area}
\end{equation}
with $\theta=\sqrt{s/a(a+1)}=1/t$ by consequence of Equation~\eqref{mean}, and mean area $s$. Let $Y$ be a random variable representing the edge length of a facet in the ridge network. If $Y$ scales as $\sqrt{X}$, the consequent distribution of $Y$ is a gamma distribution,
\begin{equation}\label{gamma}
f_Y(y)=\frac{1}{\theta\Gamma(a)}\bigg(\frac{y}{\theta}\bigg)^{a-1}e^{-y/\theta},
\end{equation}
with $\theta$ the scale and $a$ the shape parameter, alluded to in our discussion of breakup rates, and with mean edge length $a\theta$. The distributions of facet area and edge length provided by Equations~\eqref{area} \&~\eqref{gamma} allow us to formulate an expression for the typical total crease length as a function of $t$, in tandem with the evolution of mean area $s(t)$. First, we briefly restate the key empirical result of Ref.~\cite{gottesman2018state} to which we will compare our model. The total crease length $\ell$ was found to vary according to a logarithm of the number of crumpling and unfolding repetitions $n$:
\begin{equation}
    \ell_{\text{empir.}} \equiv \ell(n,\tilde{\Delta}) = c_1(1-\tilde{\Delta})\log\Bigg(1+\frac{c_2n}{\tilde{\Delta}}\Bigg), \label{eqn_of_state}
\end{equation}
with $\tilde{\Delta}$ the compaction ratio, and $c_1$ and $c_2$ fitting parameters. A striking property of this model is its implication that the rate at which new damage accumulates, as measured by added crease length per crumpling iteration $\delta{\ell}_{\text{empir.}}\equiv\partial{\ell}/\partial{n}$, is independent of the details of the sheet's preparation:
\begin{equation}
\delta\ell_{\text{empir.}} = \frac{c_1c_2\big(1-\tilde\Delta\big)}{\tilde\Delta}\exp\bigg(-\frac{\ell}{c_1(1-\tilde\Delta)}\bigg). \label{eqn_of_state_deriv}
\end{equation}
We observe from Equation~\eqref{eqn_of_state_deriv} that the added crease length $\delta\ell_{\text{empir.}}$ is
uniquely determined by a sheet's instantaneous state $(\ell, \tilde{\Delta})$; moreover, the model is independent of the details of the crease network, such as the spatial homogeneity of damage across the sheet. The fitting parameters $c_1$ and $c_2$ are universal to all values of $n$ and $\tilde{\Delta}$. The facet segmentation of each crease pattern provides a second measurable quantity, $d$, equal to the sum of all interior perimeters of facets; i.e. the total length of all edges shared between two facets. We expect $d$ and $\ell$ to be proportional, with differences arising due to incomplete scarring around facet perimeters as regions of the sheet restore elastically, particularly for mild compression. We find that $(1-\tilde{\Delta})d$ accomplishes the desired proportionality, and define $\ell_{\text{meas.}}\equiv(1-\tilde{\Delta})d_{\text{meas.}}$ to be the measured total crease length obtained from our segmented data. Next, working with the moments of our derived facet area and edge length distributions, we can estimate $d$ analytically as the average length of an edge, $a\theta$, times the average number of edges. The latter may be expressed as the average number of facets, or the total sheet area divided by the typical facet area $s$, times the number of edges per facet $n_e$, halved to account for shared edges, which yields
\begin{equation}
    d_{\text{model}} = \frac{a\theta(t)}{s(t)} \times\frac{n_e}{2} = \frac{n_e t}{2(a+1)}. \label{perim}
\end{equation}
Thus, we obtain that:
\begin{equation}
    \ell_{\text{model}}^{(t)} \equiv \ell(t,\tilde{\Delta}) = (1-\tilde{\Delta})\frac{n_e t}{2(a+1)}. \label{eqn_of_state_model}
\end{equation}
Here, the superscript $(t)$ denotes the explicit dependence of $\ell_{\text{model}}$ on $t$; in a following subsection, we develop a connection between $t$ and $n$ that allows $\ell_{\text{model}}$ to be expressed in terms of $n$ and $\tilde{\Delta}$, mirroring Equation~\eqref{eqn_of_state}. A fit of Equation~\eqref{eqn_of_state_model} to the measured length $\ell_{\text{meas.}}$ reveals a value of $n_e\approx 4.2$ as shown in Fig.~\ref{log_model_compare}b, which suggests an average of 4--5 sides per facet. Finally, Fig.~\ref{log_model_compare}c demonstrates the agreement between $\ell_{\text{model}}^{(t)}$ of Equation~\eqref{eqn_of_state_model}, and $\ell_{\text{empir.}}$ of Equation~\eqref{eqn_of_state}.

\subsection*{\label{subsec:Numerical evidence for the insensitivity to initial preparation}Numerical evidence for the insensitivity to initial preparation\protect}
\begin{figure}
\centering
\includegraphics[width=8.6cm]{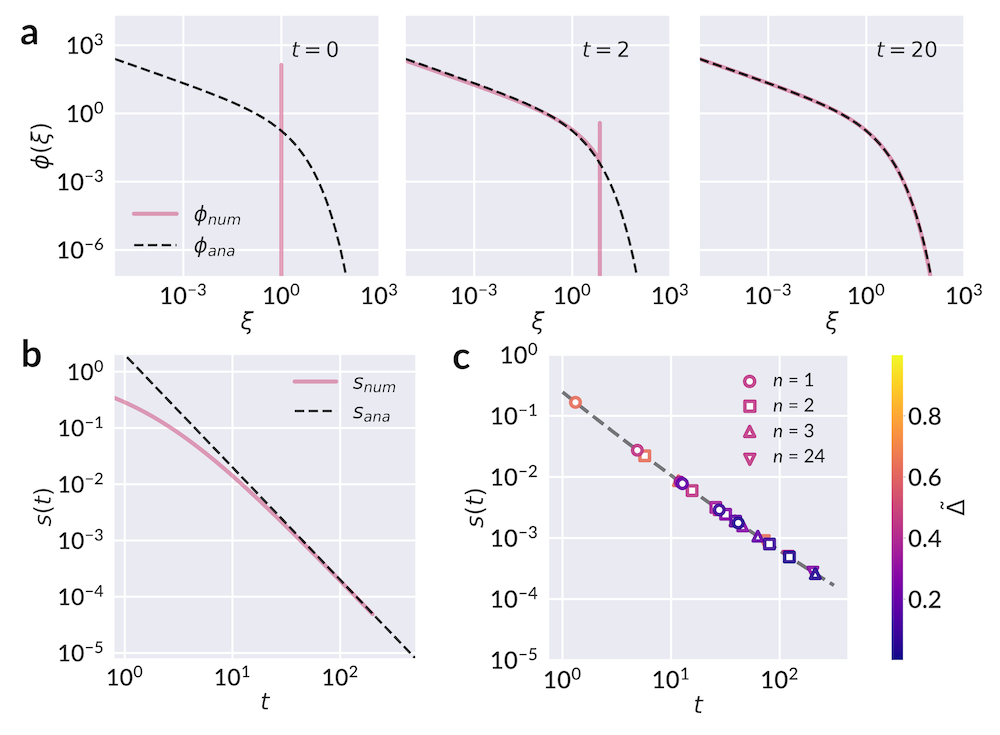}
\caption{\textbf{Numerical validation of the analytical solution to Equation~\eqref{frag_eqn}}. \textbf{a} Selected snapshots of the numerically calculated $\phi_\textit{num}(\xi)$ with initial condition $c(x,0)=\delta{(x-1)}$ and with $a=1$, revealing a rapid convergence to the steady state distribution. The dashed line corresponds to the analytical form of Equation~\eqref{phi} valid at large $t$. \textbf{b} The corresponding evolution of mean area $s(t)$, with the analytical solution at large $t$ given by Equation~\eqref{mean} shown by the dashed line. \textbf{c} The mean area of the experimental samples as a function of $t$ computed from Equation~\eqref{mean} (scattered points). The dashed line corresponds to Equation~\eqref{mean} with $a(t)$ as given by Equation~\eqref{at_eqn}. Marker colors correspond to different values of $\tilde\Delta$, as indicated by the colorbar.}
\label{num}
\end{figure}

Now that the connection between the statistical model of facet area and total crease length has been presented, we briefly note on the insight that may be gained by additionally solving Equation\eqref{frag_eqn} numerically. A numerical integration scheme is implemented using second-order composite trapezoid rule for discretization in $x$, and second-order implicit multi-step discretization in $t$. The sample numerical result in Fig.~\ref{num} reveals a rapid convergence to the steady state analytical solution given by Equations~\eqref{phi} \&~\eqref{mean}, and thereby relative insensitivity to the initial state. To demonstrate the significance of this behavior, we reiterate the observed history independence of total crease length. As discussed in Ref.~\cite{gottesman2018state}, sheets with different loading histories---one hand-crumpled and another deliberately folded along straight lines---yet nearly equal total crease lengths exhibited the same subsequent accumulation of damage when subjected to the protocol of Fig.~\ref{sample}a. Such sheets had clearly distinct initial facet area distributions: The facet areas of the deliberately folded sheet were sharply peaked near two different values, while those of the hand-crumpled sheet were broadly distributed. Thus, signatures of initial preparation appear to be quickly eclipsed by the strong attractor of the crumpled state, echoed in the rapid convergence to steady state seen numerically.

Thus far, we have established that an estimate of total crease length constructed from moments of the derived facet area and ridge length distributions shows consistency with the logarithmic scaling of Equation~\eqref{eqn_of_state}. In the following section, we propose a simple mechanism for how the geometric incompatibility of a folded sheet and its confinement leads to further fragmentation, driving $t$ forward. This argument establishes the evolution of $t$ in accordance with $n$, and thus supplies the missing link to a physically-based model that corroborates experimental findings.

\subsection*{\label{subsec:One-dimensional model}One-dimensional model\protect}
\begin{figure}
\centering
\includegraphics[width=8.6cm]{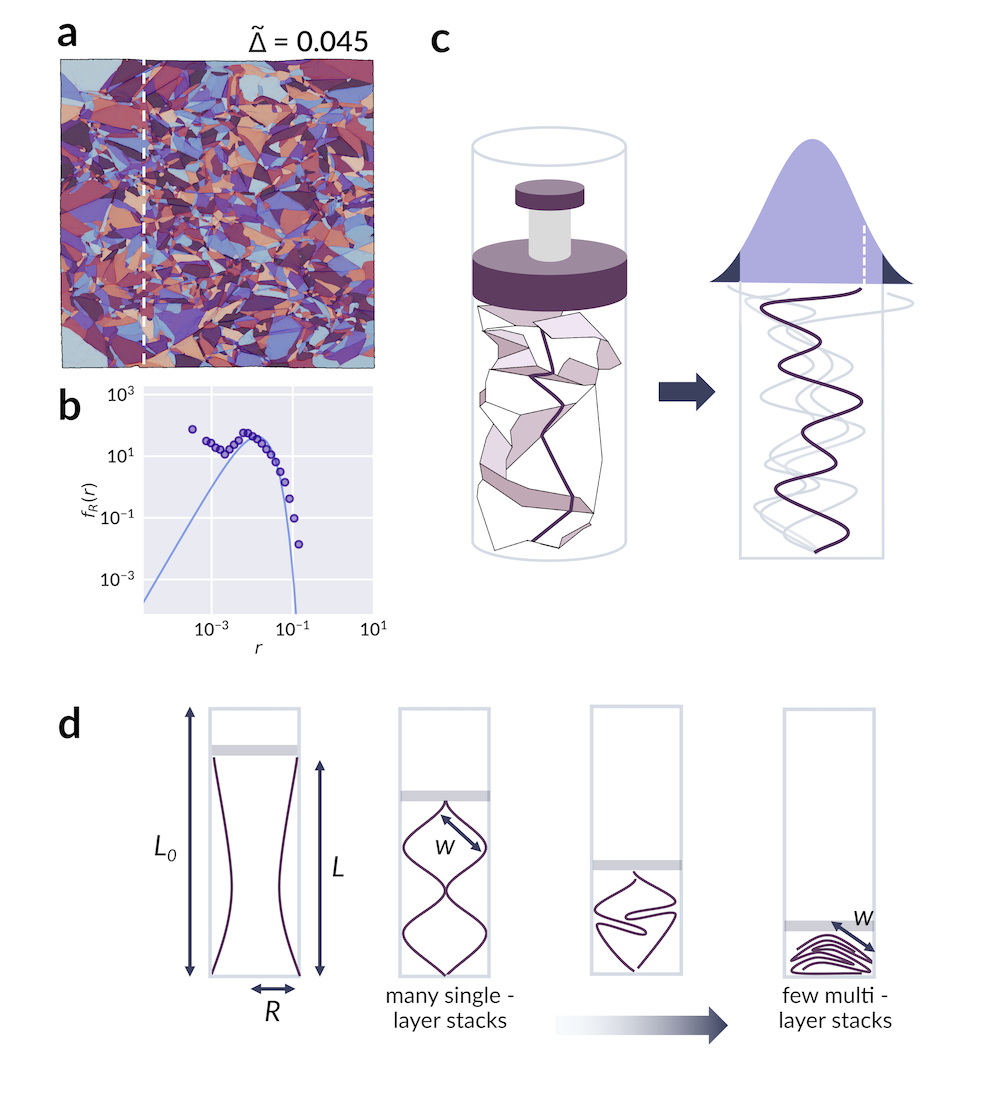}
\caption{\textbf{A folded cross-section considered as a one-dimensional random walk.} \textbf{a} A sample segmented sheet with dashed line indicating a vertical cross section. \textbf{b} The distribution of segment lengths from all such cross-sections of the sheet in \textbf{a} (filled points), with Equation~\eqref{gamma_len} plotted as a solid curve. No additional fit is performed; the value of the shape parameter $a$ which appears in Equation~\eqref{gamma_len} is uniquely determined from Equation~\eqref{at_eqn} and the best fit $\tau$ to the facet area distributions. \textbf{c} A schematic of the analogue between the folding of a one-dimensional strip in an axially confined sheet and a one-dimensional random walk whose time axis is extended vertically for clarity. The filled curve represents the distribution of the walker's final displacement, with darker shaded regions denoting the fraction of walks which lie outside a given confinement. \textbf{d} Simplified illustration of one-dimensional folding which facilitates a geometric estimate of the critical confinement $w$, further detailed in Supplementary Note 4.}
\label{random_walk}
\end{figure}

To offer an explanation for the observed logarithmic scaling, we develop a simple one-dimensional model that proposes how additional fragments may form when a crumpled sheet is re-crumpled, relying on the statistical descriptions of facet area and segment length formulated in the previous sections. Our goal can be summarized by the following two questions: (1) Given its current state and prescribed confinement, with what probability does a sheet undergo further fragmentation? (2) How does this probability relate to the continuous variables in the fragmentation model of Equation~\eqref{frag_eqn}? First, we appeal to the axial symmetry of our confinement to simplify our view of crumpling to a 1D strip of length $L_0$, as shown in Fig.~\ref{random_walk}. The strip is characterized by a sequence of folds in alternating directions which divide the strip into random segments. The lengths $r$ of the segments, which are equal to the cross-sections of the intercepted facets, are distributed according to the derived gamma distribution of Equation~\eqref{gamma}, weighted by the horizontal facet width, which increases the likelihood of a facet's occurrence within a randomly selected vertical strip. For facets of approximately $1:1$ aspect ratio, the distribution of segment length is thereby
\begin{equation}\label{gamma_len}
f_R(r)=\frac{1}{\theta\Gamma(a+1)}\bigg(\frac{r}{\theta}\bigg)^{a}e^{-r/\theta},
\end{equation}
with the average segment length given by $(a+1)\theta$. A comparison of Equation~\eqref{gamma_len} with experimental data is provided for a strongly compacted sample in Fig.~\ref{random_walk}a-b, with extended results for all samples presented in Supplementary Fig.~\ref{len_fit}.

As a preliminary step, we derive the final displacement of the strip when folded at each break, in the absence of confinement. This problem can be mapped to the displacement of a walker performing a one-dimensional random walk with gamma-distributed steps. To enforce the concept of folding, the walker's steps occur in alternating directions. The distribution $f_Z(z)$ of position $Z$ after $2k$ steps accurate for all $k$ is derived in full in Supplementary Note 2. However, the salient trends may be likewise observed by applying the central limit theorem and considering the position $Z$ valid for large $k$, or small step size, which gives:
\begin{equation}\label{displacement}
f_Z(z;\theta) = \frac{1}{\sqrt{2\pi{L_0}\theta}}\exp\Bigg(-\frac{z^2}{2L_0\theta}\Bigg),
\end{equation}
and describes a normal distribution of zero mean and variance $L_0\theta$.

\begin{figure}
\centering
\includegraphics[width=8.6cm]{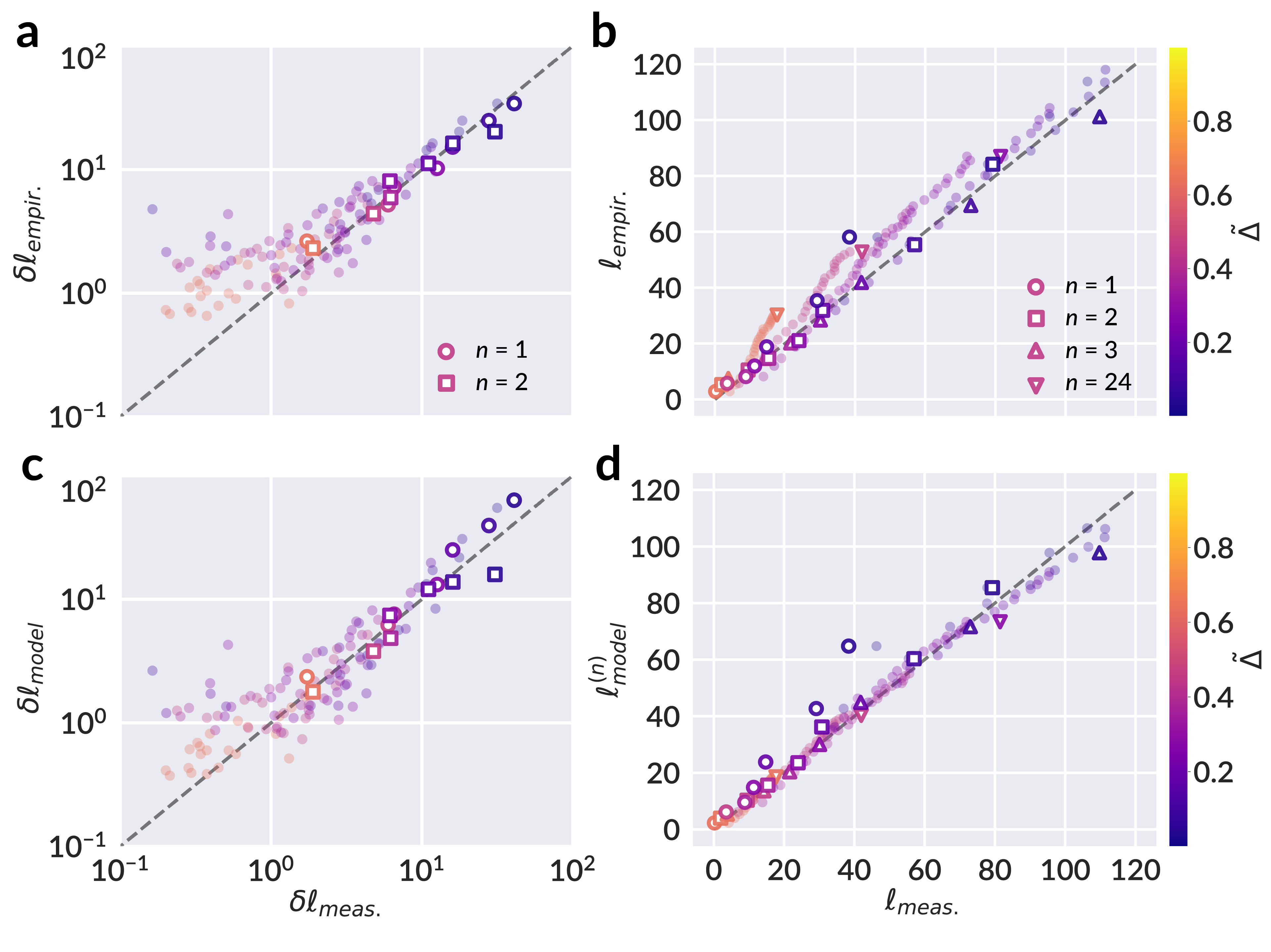}
\caption{\textbf{Validation of empirical and derived models for crease length evolution with measurement.} \textbf{a} Predicted change in total crease length $\delta\ell_{\text{empir.}}$ given by Equation~\eqref{eqn_of_state_deriv} plotted against the measured change in crease length between two successive crumples, $\delta\ell_{\text{meas.}}\equiv\ell_{\text{meas.}}^{(n)}-\ell_{\text{meas.}}^{(n-1)}$. Open markers correspond to manually segmented data consistent with prior results presented in this work, while filled circles correspond to data which was processed using the automated segmentation as detailed in the Supplementary Methods. A 1:1 reference line (dashed) is provided as a guide to the eye. \textbf{b} The total crease length $\ell_{\text{empir.}}$ given by Equation~\eqref{eqn_of_state} plotted against the measured total crease length $\ell_{\text{meas.}}$. \textbf{c} The change in total crease length $\delta\ell_{\text{model}}$ as predicted by Equation~\eqref{eqn_of_state_deriv_model}, and \textbf{d} the total crease length $\ell_{\text{model}}^{(n)}$ obtained from Equations~\eqref{eqn_of_state_model} \&~\eqref{t_of_n}, against their corresponding measured values, in direct comparison to \textbf{a} and \textbf{b}. Marker colors in all panels correspond to different values of $\tilde\Delta$, as indicated by the colorbar. We see that both the empirical and derived relations for $\delta\ell$ and $\ell$ serve as strong models of measured data, and affirm the suitability of a logarithmic relationship to describe damage evolution in this system.}
\label{diff_model_compare}
\end{figure}

If a confinement is now introduced at the locations $|z|=w$, we next ask with what likelihood the walker steps beyond this confinement. One approach to approximate this probability is to integrate Equation~\eqref{displacement} for all $|z|>w$, producing a two-sided survival function of Equation~\eqref{displacement}. Although this is not equivalent to our initial question, as intermediate steps may also have landed past $|z|>w$, it proves an acceptable estimate as the last step has the greatest variance. A more accurate calculation would be to evaluate the likelihood that a given walk escapes the confinement at any step; however, looking at the last step is useful for its simplicity in analytical form, and still captures the anticipated behavior. A comparison to the more accurate formulation is made numerically and provided in Supplementary Fig.~\ref{num_rw}. Once again, we pursue here the simpler form of the survival function valid for large $k$, and refer to Supplementary Note 2 for the exact derivation valid at all $k$. The survival function of Equation~\eqref{displacement}, $S_Z(w;\theta)=P(|Z|>w;w \geq 0)$, for a threshold confinement $w$, is given by
\begin{equation} \label{survival_fun}
S_Z(w;\theta) = 1-\text{erf}\Bigg(\frac{w}{\sqrt{2L_0\theta}}\Bigg),
\end{equation}
where $\text{erf}(z)$ is the error function. In order for walkers at $|z|>w$ to be restored within the limits of confinement, one or more of their steps must fragment, thereby increasing the number of steps taken and decreasing the overall average, which drives the evolution of fragmentation. This articulates our key claim: Considering our original, cylindrically shaped sheets as a statistical ensemble of one-dimensional random walks, we suggest that the progression of fragmentation measured by a change $dt$, over a single crumpling iteration $dn$, should be proportional to the fraction of walks in the ensemble which leave the confinement imposed at $|z|=w$: $dt/dn \sim S_Z(w;\theta)$. Equivalently, this is the likelihood that a single random walk leaves the critical confinement. We note that this resulting fragmentation rate describes an average fragmentation likelihood given only a confinement $w$ and current temporal parameter $t=1/\theta$ describing the maturity of the fragmentation process thus far; it does not enforce direct correlations between successive crumpling iterations, whereby new folds should occur preferentially along previous ones. Instead, the decrease in fragmentation rate with $n$ is encoded through the decreasing mean facet area with $t$. Moreover, while stronger correlation is expected between walks representing nearby transects of the sheet, here we consider the statistical behavior of the sheet as a whole, and account for the increased fragmentation likelihood for facets with larger horizontal extent through the weighting introduced in Equation~\eqref{gamma_len}. At present, Equation~\eqref{survival_fun} gives the likelihood that new creases will form; however, it does not yet describe how much new damage is created, for which two additional factors should be considered: (1) When the sheet is strongly confined in closely-packed layers, the layers tend to collectively fragment, as alluded to in Refs.~\cite{sultan2006statistics} and~\cite{gottesman2018state}, thus contributing a factor $p\sim1/L$ such that halving the final height doubles the number of additional ridges. (2) In the opposite limit of low compaction, facets are not in close proximity and need not behave cooperatively; thus, new damage scales linearly with the amount of compression $L_0-L$, as argued in Ref.~\cite{gottesman2018state}. With these additional considerations, we propose that the evolution of the fragmentation process with crumpling iteration behaves as
\begin{equation}\label{prog_eqn}
\delta{t} \equiv \frac{\partial{t}}{\partial{n}} =  \alpha\frac{1-\tilde{\Delta}}{\tilde{\Delta}}S_Z(w;t),
\end{equation}
where $\alpha$ is a fitted constant of proportionality. We indicate the explicit dependence on $t$ here, as $t$ and $\theta$ are inversely related. The critical width $w$ is determined by the geometry of the imposed confinement, as illustrated in Fig.~\ref{random_walk}d; a complete derivation is provided in Supplementary Note 4:
\begin{equation}
w(\tilde{\Delta}) \approx \frac{R}{\sqrt{1-\tilde{\Delta}^2}},
\end{equation}
where $R$ is the radius of the container.
By consequence of Equation~\eqref{eqn_of_state_model} we can directly relate Equation~\eqref{prog_eqn} and Equation~\eqref{eqn_of_state_deriv} as
\begin{equation}
\delta\ell_{\text{model}} = \frac{d\ell_{\text{model}}^{(t)}}{dt}\delta{t} = \frac{(1-\tilde{\Delta})n_e}{2(a+1)}\delta{t} \label{eqn_of_state_deriv_model}
\end{equation}
and obtain a fit to the proportionality constant $\alpha$. By performing an asymptotic approximation in the limit of large $t$, detailed in Supplementary Note 3, Equation~\eqref{prog_eqn} may be analytically integrated to provide a scaling relation $t(n,\tilde{\Delta})$ which bears similarity to $\ell(n,\tilde{\Delta})$ of Equation~\eqref{eqn_of_state}:
\begin{equation}
    t(n,\tilde{\Delta}) = \tilde{c}_1(1-\tilde{\Delta}^2)\log\Bigg(1+\frac{\tilde{c}_2n}{\tilde{\Delta}(1+\tilde{\Delta})}\Bigg), \label{t_of_n}
\end{equation}
where $\tilde{c}_1 = 2L_0/R^2$, $\tilde{c}_2 = \alpha R^2/L_0\sqrt{2\pi}$, and $L_0$ and $R$ are the sheet length (equivalently the confining container height) and container radius, respectively. Taken together, Equations~\eqref{eqn_of_state_model} \&~\eqref{t_of_n} thereby provide a theoretically-motivated expression $\ell_{\text{model}}^{(n)} \equiv \ell(t(n,\tilde{\Delta}),\tilde\Delta)$ based on properties of fragmentation kinetics and a simple mechanism for re-fragmentation formulated as a random walk.
Fig.~\ref{diff_model_compare} compares the agreement of the empirical relations $\delta{\ell}_{\text{empir.}}$ and $\ell_{\text{empir.}}$, as well as the derived models $\delta{\ell}_{\text{model}}$ and $\ell_{\text{model}}^{(n)}$, with the measured quantities $\delta\ell_{\text{meas.}}\equiv\ell_{\text{meas.}}^{(n)}-\ell_{\text{meas.}}^{(n-1)}$ and $\ell_{\text{meas.}}$ for various $n$. Collectively, the results of Figs.~\ref{log_model_compare} \& \ref{diff_model_compare} demonstrate clear consistency of the fragmentation model with the anticipated logarithmic growth.

\section*{\label{sec:Discussion}Discussion\protect}
By pursuing a correspondence between the crumpling of a thin sheet and a general fragmentation process, we have derived a physically-based framework for the evolution of statistical properties of intricate crumpled patterns. Equipped with theoretical models in close agreement with experimental data, we have proposed a simple model of one-dimensional folding in which further fragmentation ensues due to a geometric incompatibility between the sequence of folds and the imposed confinement, likened to a random walk exceeding a critical allowed displacement. The predicted accrual of damage, quantified by added crease length, shows strong consistency with the logarithmic model of Gottesman \textit{et al.}~\cite{gottesman2018state}, and thereby supplies a possible physical basis for the puzzling origin of logarithmic scaling in repeated crumpling experiments. Furthermore, our model explains the history independence of the logarithmic scaling, since the area distribution of the crumpled state is such a strong attractor in the fragmentation process.

The consistency of crumpling with fragmentation theory hints at the possibility of universal behavior uniting more diverse fragmenting systems. For example, the activation of defects in the fragmentation of ceramics can locally slow down subsequent fracture, and may bear semblance to the slowing of damage accumulation as a re-crumpled sheet exploits its existing folds~\cite{levy2010dynamic}. Thus, studies of crumpled systems might offer a new lens through which to interpret other complex processes. An immediate extension of this work would be a validation of the results on sheets of varied thicknesses and material parameters, as well as those prepared according to different compaction protocols. One simplifying assumption of our analysis is that fragmentation of facets is a scale invariant process over the range of areas considered; however, this assumption starts to break down particularly for large crumpling iterations $n$. The work of Ref.~\cite{lechenault2014mechanical} offers a compelling approach for identifying this limit by considering the energetic competition between bending of facets and rigid folding along existing creases, with energy cost of the former proportional to the sheet’s bending rigidity, and the latter proportional to crease stiffness. The energy balance of these competing deformations provides a characteristic length scale which varies in proportion to the sheet thickness. This improvement to the current work would strongly benefit from further studies over a range of material parameters. Length scales of folds in crumpled systems have also been studied in the context of thermally-driven dynamics, and it may thus be useful to draw possible connections to statistical mechanical models of crumpling~\cite{yllanes2019folding, bowick2001statistical, david2004statistical}. Moreover, it may be of value to explore slight generalizations of proposed functional forms introduced in this study, such as the breakup rates; this could allow variations across other experimental results to be explained, such as those arising between low and high compaction regimes~\cite{balankin2008entropic,sultan2006statistics}, thereby providing a unifying framework for such observations. 

Additionally, deeper understanding of crumpling dynamics can assist data-driven approaches to predicting damage network formation. Though machine learning methods are capable of unveiling hidden structure in complex, disordered systems~\cite{cubuk2015identifying, carrasquilla2017machine}, prior work has demonstrated the importance of preserving physical properties in making faithful predictions: for example, preserving vertex angle constraints in synthetic fold patterns to assist the task of ridge network reconstruction in crumpled sheets~\cite{hoffmann2019machine}. In addition to encoding physical rules implicitly through data, future machine learning approaches may explicitly enforce constraints such as facet area and crease length statistics in predicting ridge network evolution. Strategies which couple detailed spatial data with coarse-grained theoretical insight could thus enable more comprehensive predictions of crumpling dynamics in future studies.

\section*{\label{sec:Methods}Methods\protect}
The data analyzed in this work was collected for the study of Ref.~\cite{gottesman2018state}; here, we briefly summarize the experimental protocol for reference. \SI{10}{\centi\metre} $\times$ \SI{10}{\centi\metre} Mylar sheets are rolled into a \SI{3}{\centi\metre} diameter cylindrical container and compressed uniaxially to a specified compaction ratio $\tilde\Delta=L/L_0$, the ratio of final to initial height, with $L_0=$\SI{10}{\centi\metre} (Fig.~\ref{sample}a). The resulting ridge network inscribed on each sheet is extracted by carefully unfolding and scanning the sheet using a custom laser profilometer, which produces a height map of the sheet. A two-dimensional map of mean curvature is determined from the spatial gradients of the height profile; sharp peaks in curvature mark the signature of a ridge (Fig.~\ref{sample}b). Successive re-crumpling and scanning of a single sheet is performed $n$ times up to $n=24$. Individual facets, characterized as contiguous regions of near-zero curvature, are delineated as shown in Fig.~\ref{sample}c. Due to noise and artifacts in data collection, not all facets are completely enclosed by a contour of ridges; breaks along a ridge, or smoothing out and softening of ridges, occur inevitably during re-crumpling and unfolding. Automated methods of crease detection and facet labeling were initially tested to perform the segmentation; however, these methods proved sensitive to noise and thus were prone to over-fragmenting the sheets. Each sample presented and analyzed in this work was digitally labeled by hand. Additional details of automated segmentation are provided in the Supplementary Methods, and resulting segmentations and facet area distributions are shown in Supplementary Figs.~\ref{data_watershed} \&~\ref{fit_watershed}. With manual segmentation, care was taken to identify not only the dominant lines of each pattern as seen in the examples, but also the less pronounced softer scarring. The segmentation was performed for sheets after iterations $n=1,2$ and $3$ at seven different compaction ratios: $\tilde{\Delta} = 0.63, 0.45, 0.36, 0.27, 0.18, 0.09$ and $0.045$. Each series of successive crumples was compared across all iterations $n$ for consistency, to ensure that labeled facets from earlier iterations persist in later ones. Samples with $\tilde{\Delta} = 0.63, 0.45$, and $0.27$ were also labeled after $n=24$ crumples, for a total of $24$ samples overall. We acknowledge that samples at $n=24$ are more prone to missing detail as older scarring is obscured by newer ridges, but are nonetheless valuable to the study. The results of manual segmentation and corresponding facet area distributions are provided in Supplementary Figs.~\ref{data} \&~\ref{fit}.

\begin{acknowledgments}
This work was supported by the National Science Foundation under Grant No.~DMR-2011754. J.A.\ acknowledges support from the National Science Foundation under Grant No.~DGE-1745303. C.H.R.~was partially supported by the Applied Mathematics Program of the U.S.\ DOE Office of Science Advanced Scientific Computing Research under Contract No.~DE-AC02-05CH11231.
\end{acknowledgments}

\noindent \textbf{Author contributions} S.M.R. and C.H.R. conceived the study and supervised the project. L.M.L. performed new experiments. J.A. performed the image segmentation. J.A. and C.H.R. carried out analytical derivations and numerical validation. J.A. and C.H.R. wrote the paper with input from all authors.

\noindent \textbf{Competing interests} The authors declare no competing interests.

\noindent \textbf{Data availability} No new experimental data was produced for this study; all data analyzed was previously collected and reported in Ref.~\cite{gottesman2018state}. The subset of data from Ref.~\cite{gottesman2018state} used in this article is provided in post-processed form with our analysis codes.

\noindent \textbf{Code availability} The data processing and analysis codes are available on GitHub at \href{https://github.com/jandrejevic12/fragmentation_model}{https://github.com/jandrejevic12/fragmentation\_\,model}~\cite{jandrejevic12}.

\bibliography{main}
\clearpage
\renewcommand{\figurename}{Supplementary Figure}

\setcounter{section}{0}
\setcounter{equation}{0}
\setcounter{figure}{0}
\setcounter{table}{0}
\setcounter{page}{1}
\makeatletter
\renewcommand\theHequation{S\arabic{equation}}
\renewcommand\theHfigure{S\arabic{figure}}

\onecolumngrid
\begin{center}
    \large\textbf{A model for the fragmentation kinetics of crumpled thin sheets} \\
    \vspace{\baselineskip}
    \normalsize Jovana Andrejevic,\textsuperscript{1} Lisa M.~Lee,\textsuperscript{1} Shmuel M.~Rubinstein,\textsuperscript{2} and Chris H.~Rycroft\textsuperscript{1,3} \\
    \textsuperscript{1}\textit{John A.~Paulson School of Engineering and Applied Sciences,} \\
    \textit{Harvard University, Cambridge, MA 02138, USA} \\
    \textsuperscript{2}\textit{The Racah Institute of Physics, The Hebrew University of Jerusalem, Jerusalem 91904, Israel} \\
    \textsuperscript{3}\textit{Computational Research Division, Lawrence Berkeley Laboratory, Berkeley, CA 94720, USA}
\end{center}

\section*{Supplementary Figure 1}
\begin{figure}[htb!]
\centering
\includegraphics[width=\textwidth]{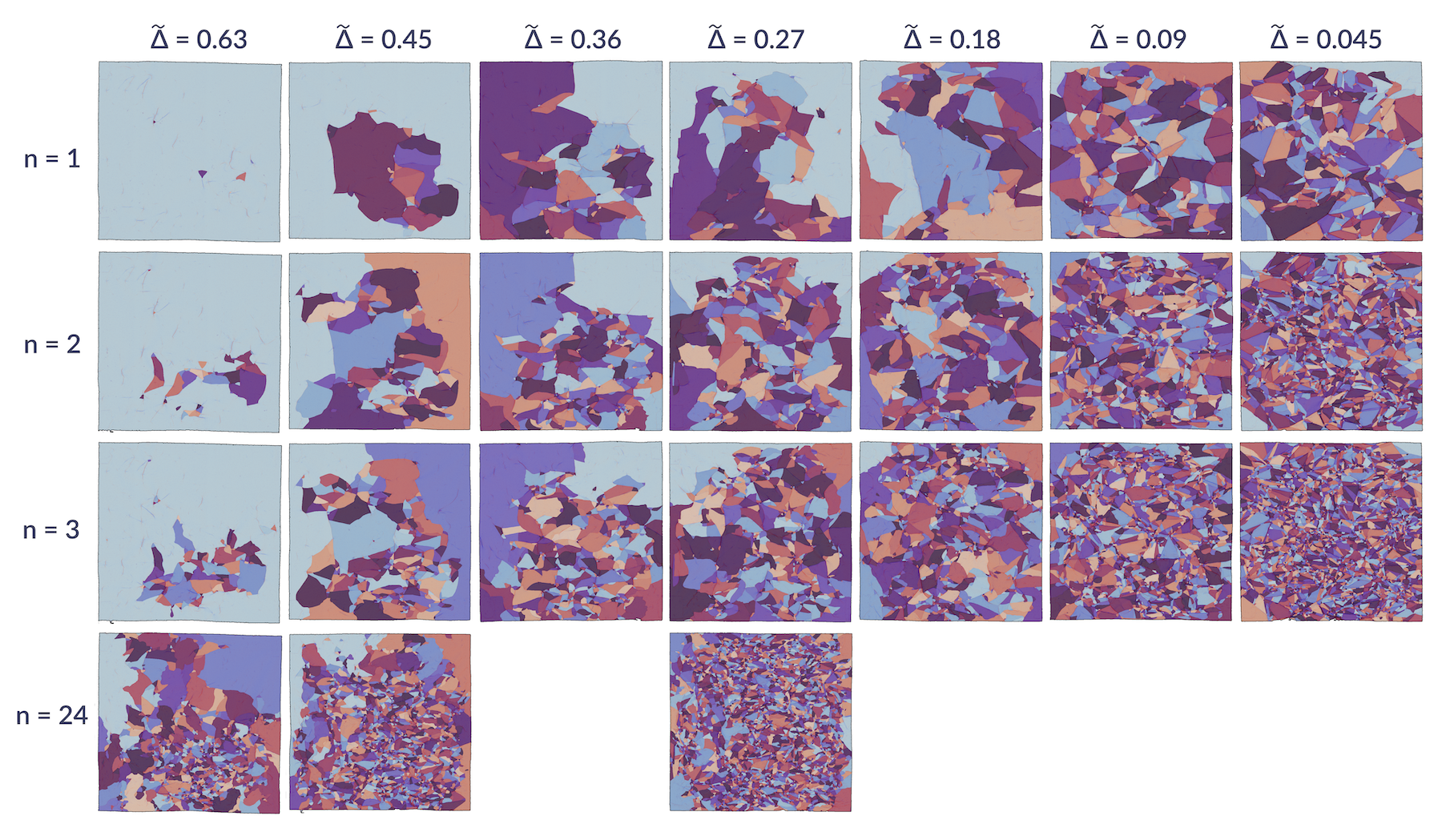}
\caption{\textbf{Manual facet segmentation.} Scans of crumpled sheets manually segmented into individual facets delineated by creases. Each column features a single sheet crumpled repeatedly $n$ times to a specified compaction ratio $\tilde{\Delta} = L/L_0$. Random coloring is used for visual distinction between facets.}
\label{data}
\end{figure}

\clearpage
\section*{Supplementary Figure 2}
\begin{figure}[htb!]
\centering
\includegraphics[width=\textwidth]{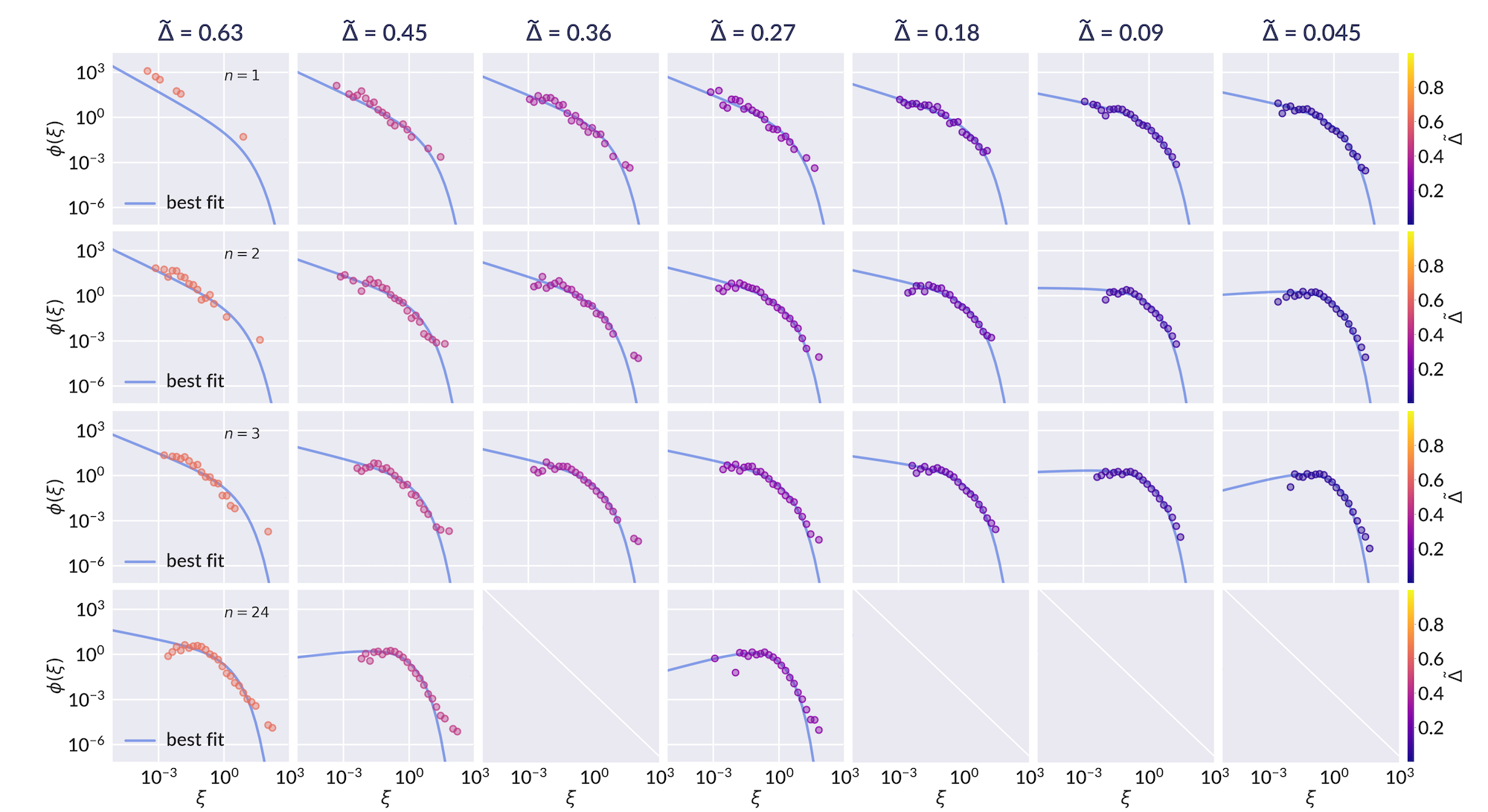}
\caption{\textbf{Facet area distributions from manual segmentation.} Distributions of facet areas $x$ normalized by the mean area $s$ respectively for each sample; $\xi=x/s$. The solid line shows the best fit to Supplementary Equation~\eqref{phi_supp}, where the parameter $a$ is calculated from the relation $a(t)=\sqrt{t/\tau}$ presented in the main text with single fitting parameter $\tau$ across the entire dataset. Marker colors correspond to different values of $\tilde\Delta$, as indicated by the colorbar.}
\label{fit}
\end{figure}

\clearpage
\section*{Supplementary Figure 3}
\begin{figure}[htb!]
\centering
\includegraphics[width=0.85\textwidth]{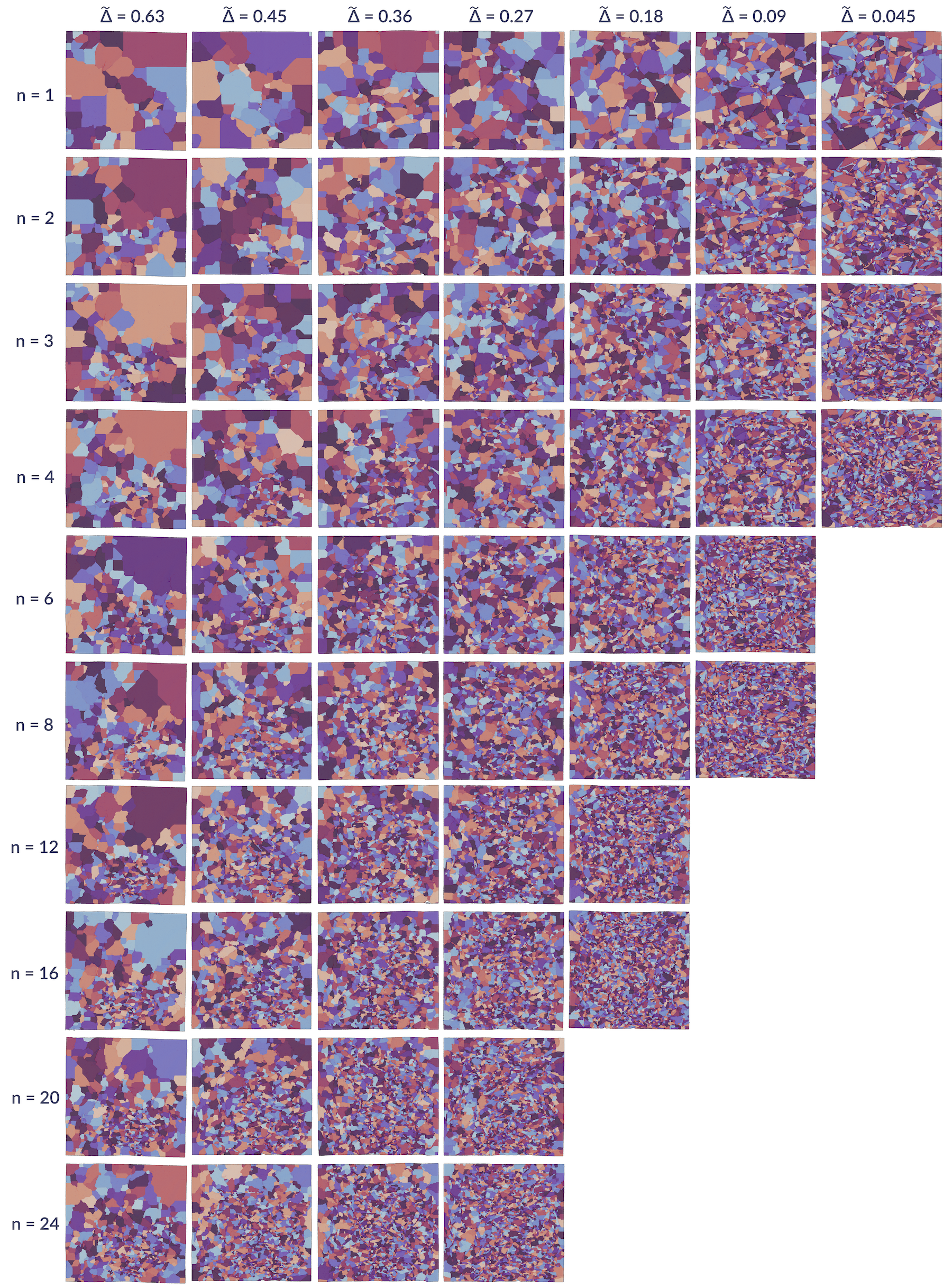}
\caption{\textbf{Automated facet segmentation.} Results of automated segmentation, shown for the same experimental samples as Supplementary Fig.~\ref{data}, and including a larger representation of crumpling iterations. Random coloring is used for visual distinction between facets.}
\label{data_watershed}
\end{figure}

\clearpage
\section*{Supplementary Figure 4}
\begin{figure}[htb!]
\centering
\includegraphics[width=0.85\textwidth]{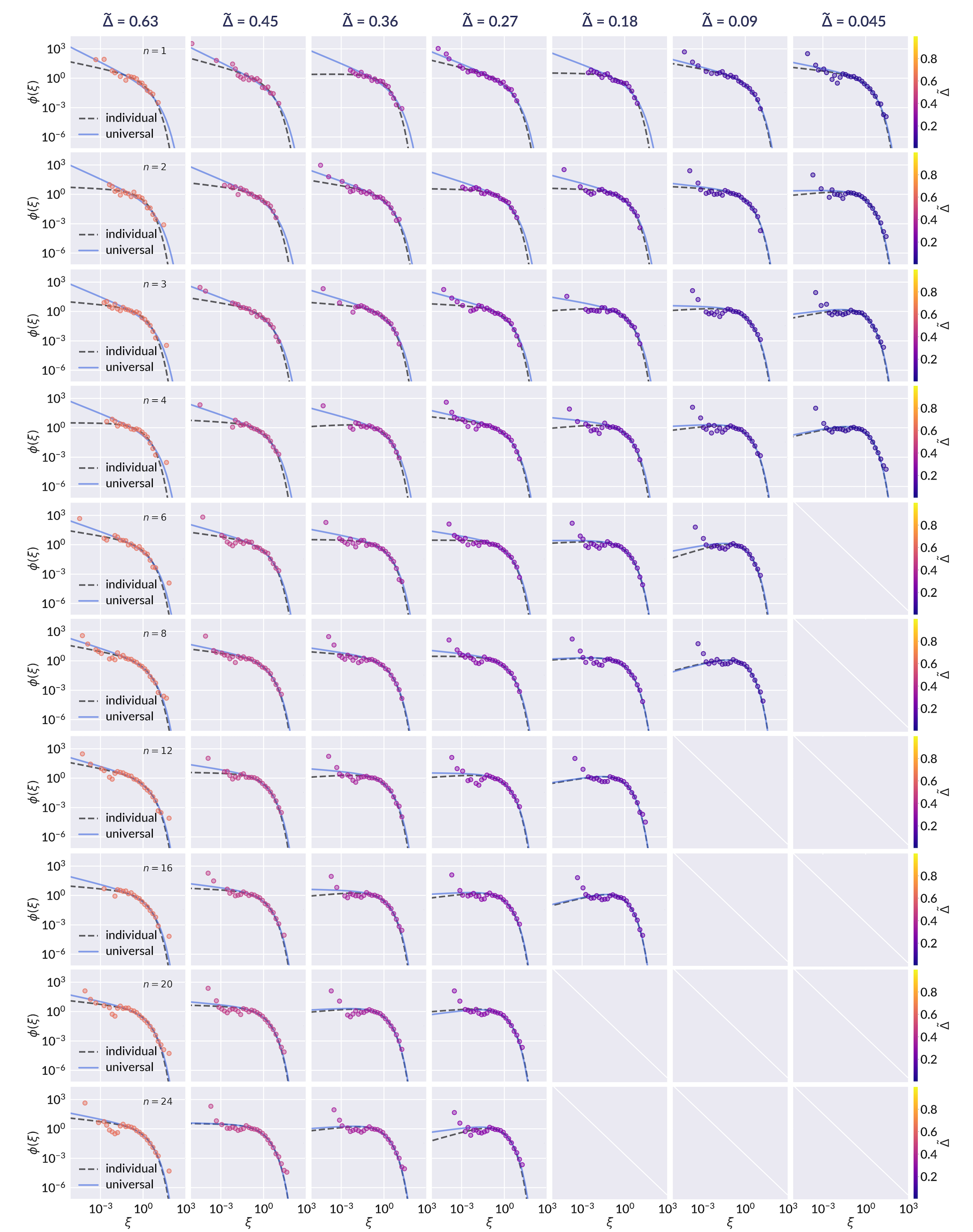}
\caption{\textbf{Facet area distributions from automated segmentation.} Corresponding distributions of normalized facet areas $\xi$ for the data in Supplementary Fig.~\ref{data_watershed}. The two accompanying curves show the best individual fit to Supplementary Equation~\eqref{phi_supp} with fitting parameter $a$ (dashed line), and the curve obtained via the relation $a = \sqrt{t/\tau}$ with universal parameter $\tau$ fit from the manually segmented data (solid line). Marker colors correspond to different values of $\tilde\Delta$, as indicated by the colorbar. Due to the automated method's sensitivity to artifacts in crease detection, we see that weakly crumpled sheets tend to be over-partitioned, and the best fit $a$ deviates from the predicted trend with $t$. However, denser crease networks demonstrate improved agreement between the individually fitted and predicted values of $a$.}\vspace{-1em}
\label{fit_watershed}
\end{figure}

\clearpage
\section*{Supplementary Figure 5}
\begin{figure}[htb!]
\centering
\includegraphics[width=\textwidth]{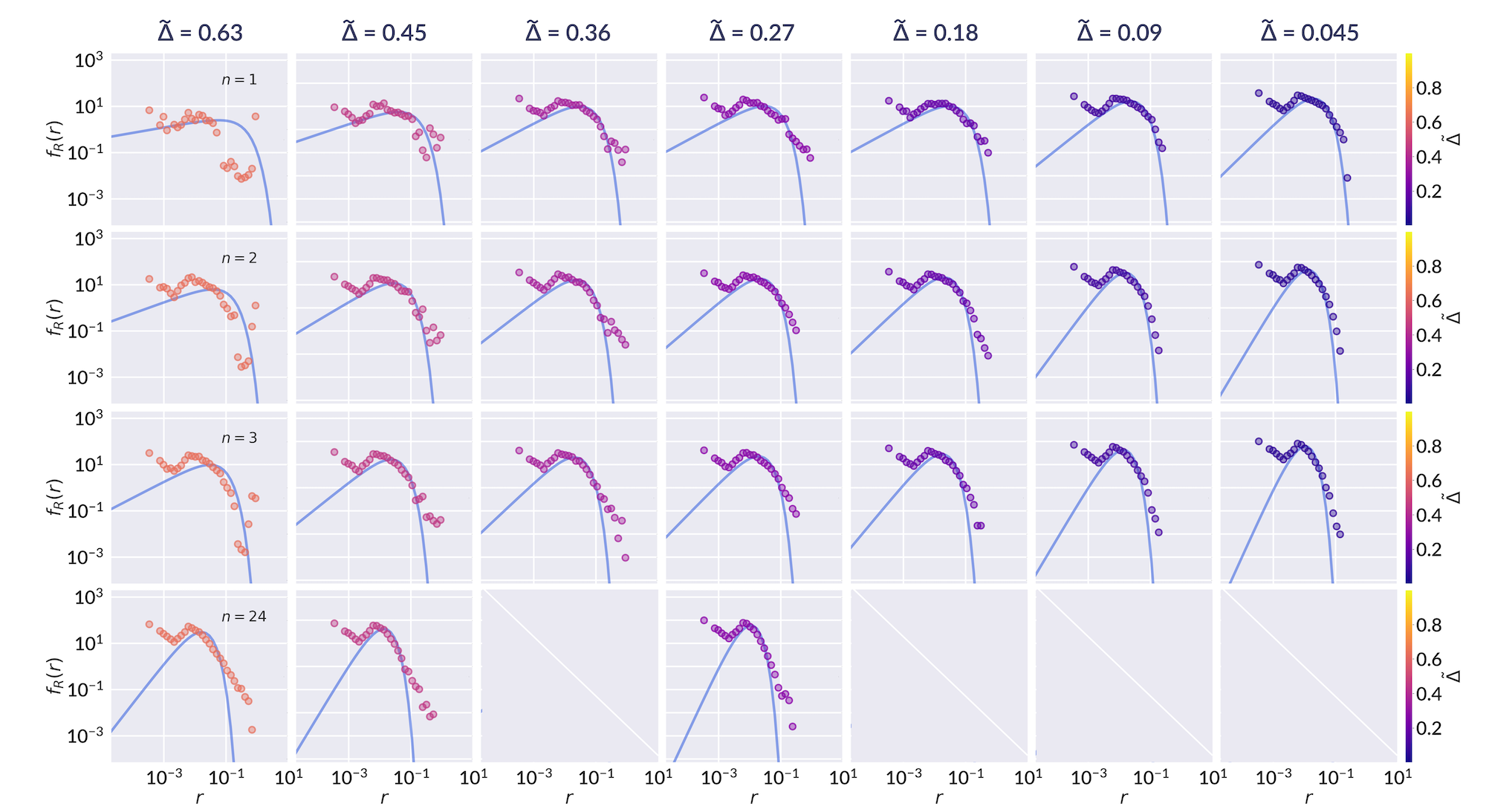}
\caption{\textbf{Facet length distributions from manual segmentation.} The distribution of facet lengths obtained by considering all vertical cross-sections of the facet patterns in Supplementary Fig.~\ref{data} (filled points), with the analytically derived distribution of Equation~(15) of the main text plotted as a solid curve. Marker colors correspond to different values of $\tilde\Delta$, as indicated by the colorbar. Note that no additional fit is performed; the shape parameter $a$ which appears in Equation~(15) of the main text is the same as that obtained for the facet area distributions with a single fitting parameter $\tau$, as detailed in the main text.}
\label{len_fit}
\end{figure}

\clearpage
\section*{Supplementary Figure 6}
\begin{figure}[htb!]
\centering
\includegraphics[width=0.85\textwidth]{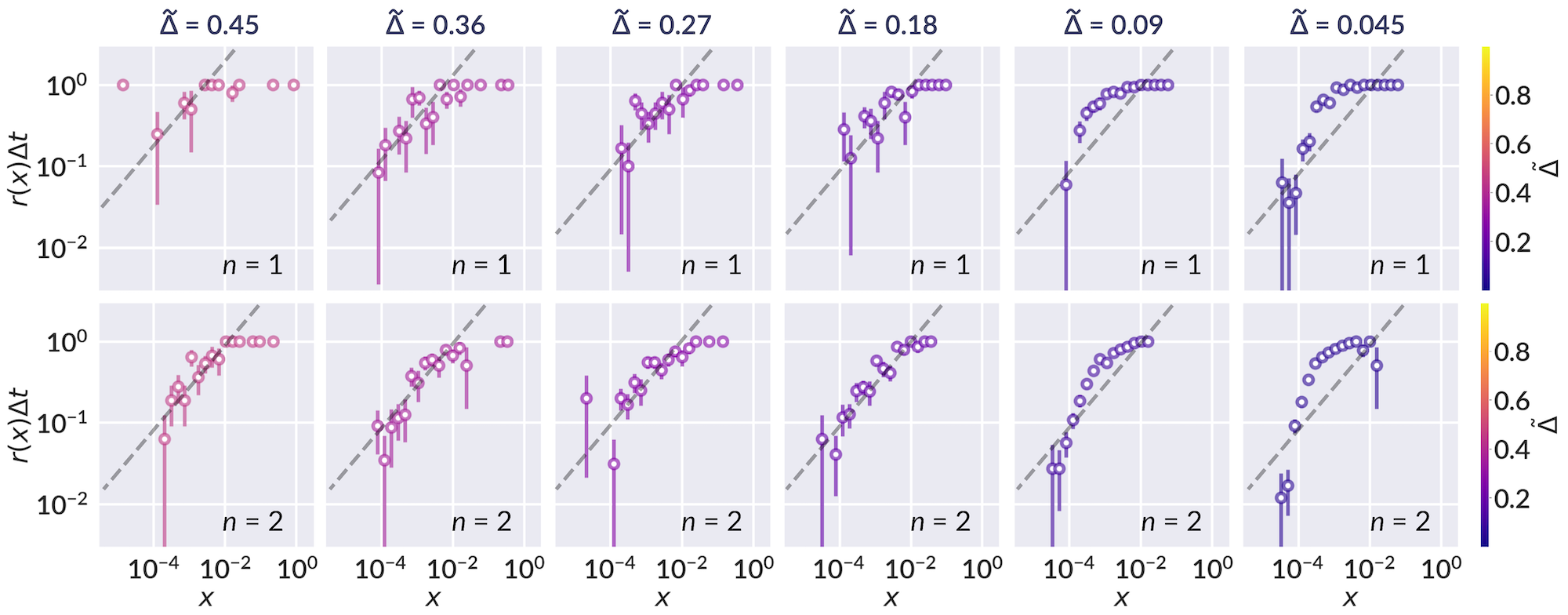}
\caption{\textbf{Estimation of overall breakup rate $\mathbf{r(x)}$.} For $n=1$ (top row), the fraction of facets present at crumpling iteration $n=1$ which fragment by $n=2$, as a function of initial area $x$, for samples with $\tilde{\Delta}=0.45,0.36,0.27,0.18,0.09,$ and $0.045$ (across). For $n=2$ (bottom row), the fraction of all facets present at crumpling iteration $n=2$ which fragment by $n=3$ for the same samples. The sample with $\tilde{\Delta}=0.63$ had too few facets to form a sufficient representative sample. Error bars denote the standard deviation of the fragmentation probability if the fragmentation of each facet is regarded as a Bernoulli trial, with the fraction of fragmented facets taken as the success probability within each histogram bin. The dashed line corresponds to $\sqrt{x}$. Marker colors correspond to different values of $\tilde\Delta$, as indicated by the colorbar. As noted in the text, samples at small values of $\tilde{\Delta}$, or high compaction, likely undergo a succession of fragmentation events between $n=i$ and $n=i+1$, and are thus poorer indicators of the statistics of single breakup events. Samples at large values of $\tilde{\Delta}$ are more likely resolve single breakup events, but have a lower population of facets from which to build the distribution. The choice of overall breakup rate $r(x)=x^{1/2}$ was motivated both by the stronger power law behavior at high $\tilde{\Delta}$, as well as its tractability in our analytical model.}
\label{rx_full}
\end{figure}

\section*{Supplementary Figure 7}
\begin{figure}[htb!]
\centering
\includegraphics[width=0.85\textwidth]{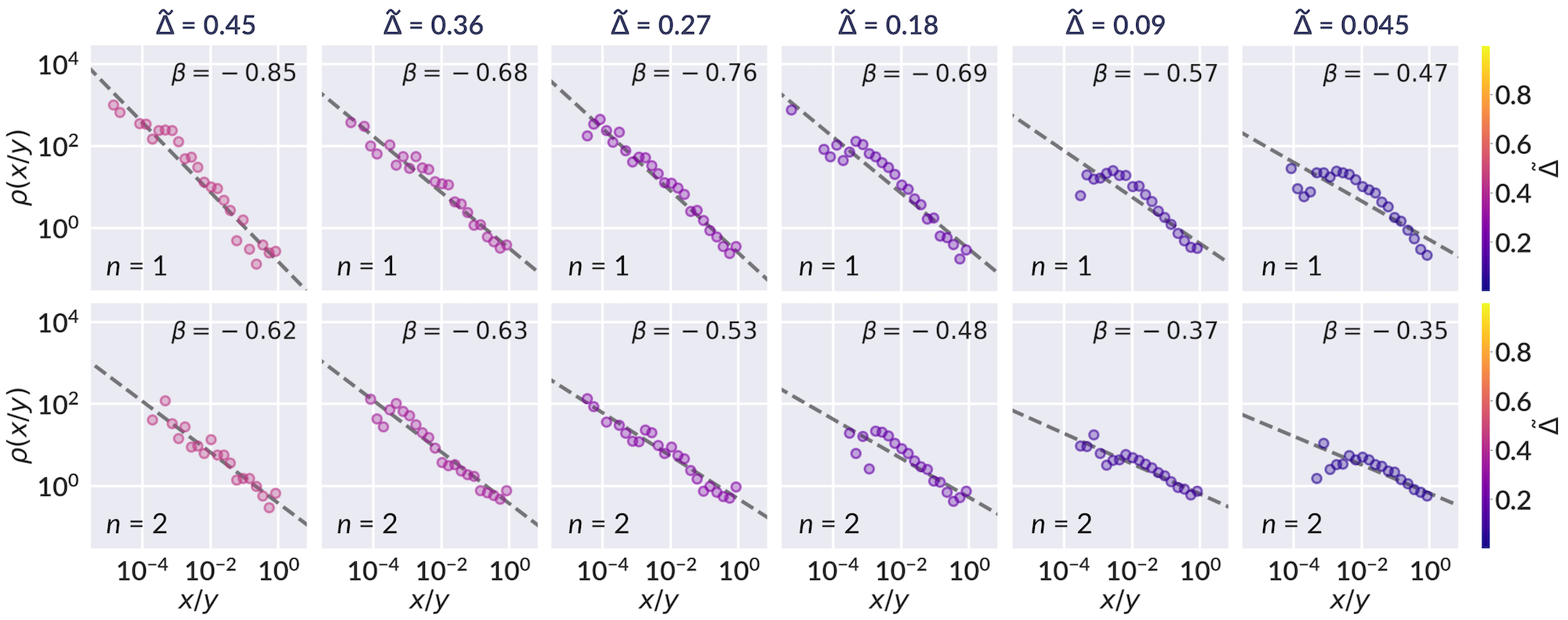}
\caption{\textbf{Estimation of conditional breakup probability $\mathbf{f(x|y)}$.} For $n=1$ (top row), the probability density function $\rho(x/y)$ of facet areas $x$ present in crumpling iteration $n=2$ normalized by their parent facet's area $y$ from $n=1$, for samples with $\tilde{\Delta}=0.45,0.36,0.27,0.18,0.09,$ and $0.045$ (across). For $n=2$ (bottom row), $\rho(x/y)$ of facet areas in $n=3$ normalized by their parent facet's area from $n=2$. Marker colors correspond to different values of $\tilde\Delta$, as indicated by the colorbar. The sample with $\tilde{\Delta}=0.63$ had too few facets to form a sufficient representative sample and is excluded here. As noted in Supplementary Fig.~\ref{rx_full}, samples at small values of $\tilde{\Delta}$ likely undergo a succession of fragmentation events between crumples, and thus their distributions resemble the more mature facet distributions observed at later $n$, as in Supplementary Fig.~\ref{fit}.}
\label{fxy_full}
\end{figure}

\clearpage
\section*{Supplementary Figure 8}
\begin{figure*}[htb!]
\centering
\includegraphics[width=0.9\textwidth]{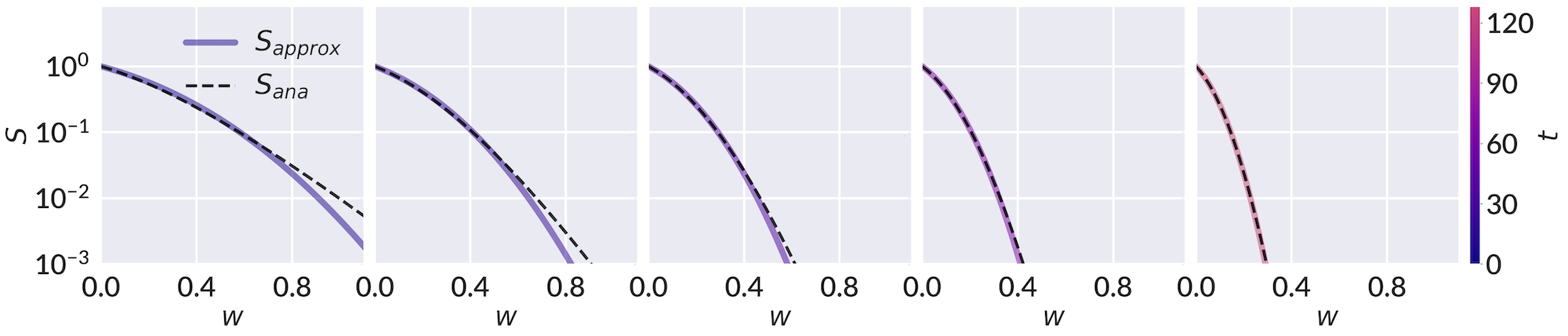}
\caption{\textbf{Convergence of asymptotic approximation for survival function $\mathbf{S_Z(w;}\bm{\theta)}$.} Plot of Supplementary Equation~\eqref{survival_fun_exact} (dashed line) against the asymptotic approximation of Supplementary Equation~\eqref{survival_fun_approx} valid at large number of steps $k$ (solid line), at $2k=4, 8, 16, 32,$ and $64$ steps from left to right, respectively. The shape parameter $a=1$ for all cases, and $t$ is appropriately determined from the relation $t=2k(a+1)/L_0$. Curves of the asymptotic approximation are colored by the value of $t$, as indicated by the colorbar. The approximation shows increasing agreement with the exact solution for larger $k$, as anticipated.}
\label{num_rw_approx}
\end{figure*}

\section*{Supplementary Figure 9}
\begin{figure*}[htb!]
\centering
\includegraphics[width=\textwidth]{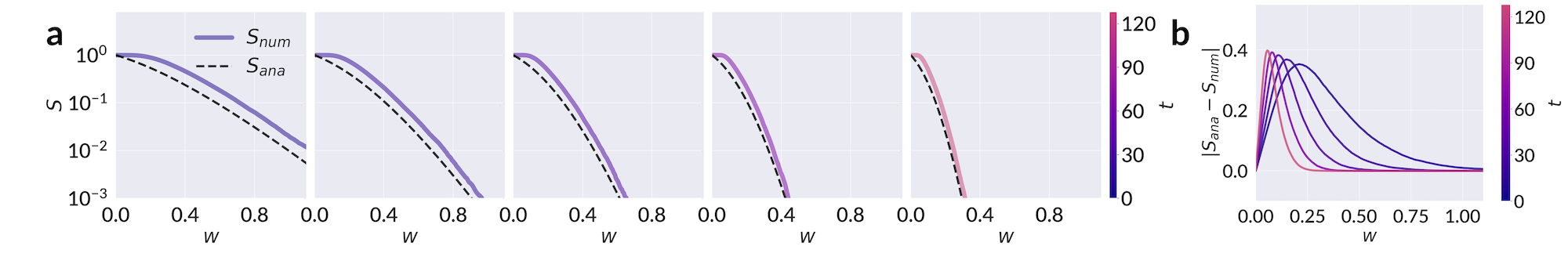}
\caption{\textbf{Numerical simulation of random walks and exact escape probability compared to analytical approximation.} \textbf{a} Plot of Supplementary Equation~\eqref{survival_fun_exact}, which presents an analytical estimate of the fraction of escaped random walks looking only at final displacements (dashed line), compared against the more accurate result obtained by assessing all intermediate displacements through numerical simulation (solid line). For the latter, simulations of $50,000$ random walks, with gamma-distributed steps sampled according to Supplementary Equation~\eqref{displacement_exact} with shape parameter $a=1$, are performed with $2k=4, 8, 16, 32,$ and $64$ steps, and $t$ appropriately determined from the relation $t=2k(a+1)/L_0$. Curves corresponding to numerical simulation are colored by the value of $t$, as indicated by the colorbar. The analytical approximation systematically underestimates the number of escaped walks by approximately a constant factor; thus, $S_Z(w;a,\theta)$ remains proportional to the change in $t$ over a crumpling iteration in both the analytically approximate and numerically calculated forms, allowing variation in the constant of proportionality. \textbf{b} The error between each pair of curves presented in \textbf{a}.}
\label{num_rw}
\end{figure*}

\clearpage
\section*{\label{sec:Complete Data}Supplementary Methods\protect}
\noindent\textbf{Manual segmentation.}
As noted in the main text, the final segmentation of all collected crease networks was performed by hand. These segmentations and the distributions of scaled facet areas are provided in Supplementary Figs.~\ref{data} \&~\ref{fit}, respectively. We recognize in the Methods section of the main text that creases can soften over repeated crumples, and the result of unfolding and scanning between crumples could possibly contribute to the appearance of healing. To ensure that the crease patterns studied suitably fit the framework of a fragmentation model, we perform a simple analysis to affirm that healing is indeed very minimal. We make a quantitative prediction about the extent of healing by overlaying two manually segmented crease patterns from successive crumples $n-1$ and $n$ of the same sheet, and measuring the length of creases present in $n-1$ which do not appear in $n$. We observe that the percent of healed creases make up less than $5\%$ of any crease pattern; moreover, the fraction is typically within $1\%$ for moderately to highly dense patterns. Thus, we conclude that healing is a small effect which does not greatly impact the dominant trends in the data.\\*\\*
\noindent\textbf{Watershed segmentation.} Prior to manual segmentation, an automated method using the watershed algorithm was initially tested. To perform this method, the maps of mean curvature for each sheet were first thresholded to produce a binary image separating creases, a pixel value of 1, and background, a pixel value of 0. The distance of each background pixel to its nearest crease pixel was then computed. The negative distance, which can be regarded as a topographic surface of hills and basins, was used as the elevation map for the watershed algorithm. The basins correspond to local minima of the surface, and regions centered at each basin are flooded until all pixels are assigned a basin. We identified pixels belonging to the same basin of the elevation map as a single facet of the crease pattern. However, several concerns prompted a more careful labeling by hand. Firstly, the separation of creases from background was performed using a custom technique referred to as the Radon transform method, detailed in the Supplementary Discussion of Ref.~\cite{gottesman2018state}. This technique combines global and local thresholding to accommodate variations in the intensity (curvature) of creases; nevertheless, softening of old creases near strongly imprinted ones weakens their detection. The watershed algorithm proved sensitive to creases which scar the sheet but do not form closed contours, particularly true at low confinement (high $\tilde{\Delta}$). Thus, the algorithm over-partitions the crease network in these cases. Mitigating the effect of smaller, isolated creases and vertices by stricter thresholding also compromises the detection of smaller facets, impacting densely scarred samples at low $\tilde{\Delta}$. The results of watershed segmentation and corresponding scaled facet area distributions are presented in Supplementary Figs.~\ref{data_watershed} \&~\ref{fit_watershed}; while there is consistency with the manually segmented data, lower resolution and weaker performance for small features impacts the range over which consistent scaling is observed. Nevertheless, the automated method allows us to more easily segment a larger number of crease patterns; thus, we process a more extensive set of crumpling iterations for each compaction ratio considered.
\clearpage

\section*{\label{sec:Scaling solution to the fragmentation rate equation}Supplementary Note 1\protect}
\noindent\textbf{Scaling solution to the fragmentation rate equation.} Facet fragmentation is modeled following the theory of fragmentation kinetics outlined in~\cite{cheng1990kinetics}. Here we reproduce the derivation of a scaling solution shared among setups with similar families of breakup rates, as well as carry out the analytical steps unique to our specific choice of these rates. The linear integro-differential equation describing the evolution of concentration of facet areas $x$, $c(x,t)$, is given by:
\begin{equation} \label{frag_eqn_supp}
\frac{\partial{c}(x,t)}{\partial{t}} = \underbrace{-r(x)c(x,t)}_{\text{depletion of facets of area $x$}} + \int_x^{\infty}\underbrace{c(y,t)r(y)f(x|y)dy}_{\text{gain in facets of area $x$}},
\end{equation}
where
\begin{equation*}
\begin{aligned}
t & \qquad \text{measure of progression of fragmentation} \\
r(x) & \qquad \text{overall rate at which a facet of area $x$ breaks} \\
f(x|y) & \qquad \text{conditional probability that $x$ is produced from the breakup of $y$, $y\geq x$} \\
c(x,t) & \qquad \text{concentration of facets of area $x$}
\end{aligned}
\end{equation*}
with the scaling ansatz
\begin{equation*}
c(x,t) = \frac{1}{s^2}\phi\bigg(\frac{x}{s}\bigg),
\end{equation*}
which restricts all time dependence to a parameter $s=s(t)$ that represents the typical (mean) area, and $\phi(\xi)$ is a scaling function. The scaling function satisfies
\begin{equation*}
\begin{aligned}
\int_0^{\infty}\phi(\xi)d\xi &= 1 \\
\int_0^{\infty}\xi\phi(\xi)d\xi &= 1
\end{aligned}
\end{equation*}
such that
\begin{equation*}
\begin{aligned}
\int_0^{\infty}c(x,t)dx = \frac{1}{s(t)} & \qquad \text{(average number of facets)} \\
\int_0^{\infty}xc(x,t)dx = 1 & \qquad \text{(total area)},
\end{aligned}
\end{equation*}
which ensures conservation of area.
A common choice of $r(x)$ and $f(x|y)$ which prove analytically tractable are members of homogeneous kernels:
\begin{equation*}
\begin{aligned}
r(x) &= x^{\lambda}, \\
f(x|y) &= \frac{1}{y}b\bigg(\frac{x}{y}\bigg).
\end{aligned}
\end{equation*}
With this formulation, larger facets are more likely to split due to the higher rate given by $r(x)$ assuming $\lambda>0$. Moreover, the conditional probability must satisfy area conservation,
\begin{equation*}
\int_0^y xf(x|y)dx = y.
\end{equation*}
Plugging the scaling ansatz and general homogeneous kernels into the rate equation, and defining $\xi=x/s, \eta = y/s$ yields
\begin{equation*}
\begin{aligned}
\frac{\partial\bigg(\frac{1}{s^2}\phi(\xi)\bigg)}{\partial{t}} &= -x^{\lambda}\frac{1}{s^2}\phi(\xi) + \int_x^{\infty}\frac{1}{s^2}\phi(\eta)y^{\lambda}\frac{1}{y}b\bigg(\frac{\xi}{\eta}\bigg)dy, \\
-\frac{2}{s^3}\dot{s}\phi(\xi)+\frac{1}{s^2}\phi'(\xi)\bigg(-\frac{x}{s^2}\bigg)\dot{s} &= -s^{\lambda-2}\xi^{\lambda}\phi(\xi) + \int_\xi^{\infty}s^{\lambda-2}\phi(\eta)\eta^{\lambda-1}b\bigg(\frac{\xi}{\eta}\bigg)d\eta, \\
-\dot{s}s^{-(\lambda+1)}\Big(2\phi(\xi)+\xi\phi'(\xi)\Big) &= -\xi^{\lambda}\phi(\xi) + \int_\xi^{\infty}\phi(\eta)\eta^{\lambda-1}b\bigg(\frac{\xi}{\eta}\bigg)d\eta,
\end{aligned}
\end{equation*}
where $\dot{s} \equiv ds/dt$. By separating the dependence on $x$ and $t$ we must have that
\begin{equation*}
-\dot{s}s^{-(\lambda+1)} = \frac{-\xi^{\lambda}\phi(\xi) + \int_\xi^{\infty}\phi(\eta)\eta^{\lambda-1}b\bigg(\frac{\xi}{\eta}\bigg)d\eta}{2\phi(\xi)+\xi\phi'(\xi)} = \omega = \text{constant}
\end{equation*}
and thus have two equations
\begin{subequations}
\begin{align}
\omega\Big(2\phi(\xi)+\xi\phi'(\xi)\Big) &= -\xi^{\lambda}\phi(\xi) + \int_\xi^{\infty}\phi(\eta)\eta^{\lambda-1}b\bigg(\frac{\xi}{\eta}\bigg)d\eta \label{sep2} \\
\dot{s}s^{-(\lambda+1)} &= -\omega, \label{sep1}
\end{align}
\end{subequations}
Insight from experimental facet fragmentation data reveals a suitable form for the conditional breakup rate:
\begin{equation*}
b\bigg(\frac{x}{y}\bigg) = \bigg(\frac{\beta+2}{\beta+1}\bigg)\rho\bigg(\frac{x}{y}\bigg),
\end{equation*}
where
\begin{equation*}
\begin{aligned}
\rho\bigg(\frac{x}{y}\bigg) &= (\beta+1)\bigg(\frac{x}{y}\bigg)^{\beta} \\
\end{aligned}
\end{equation*}
with $\beta$ a free parameter and $\rho(x/y)$ the probability density function of facet areas $x$ normalized by their parent facet's area $y$ from the previous crumpling iteration. In other words, $\rho(x/y)d(x/y)$ is the probability that a facet breaks to produce a fragment that is $x/y$ of its initial area. This formulation introduces the assumption that fragmentation is a scale invariant process.

Next we demonstrate the agreement of our scaling function with the rate equation. We begin by re-expressing $\beta$ in terms of a new parameter $a$ as
\begin{equation*}
\beta = \frac{a}{2}-1.
\end{equation*}
Our proposed solution $\phi(\xi)$ takes the form
\begin{equation*}
\phi(\xi) = \frac{\lambda}{\Gamma\Big(\frac{a}{2\lambda}\Big)}G(a,\lambda)\big(G(a,\lambda)\xi\big)^{\frac{a}{2}-1}e^{-\big(G(a,\lambda)\xi\big)^\lambda},
\end{equation*}
and thus
\begin{equation*}
\begin{aligned}
\phi'(\xi) &= \frac{\lambda}{\Gamma\Big(\frac{a}{2\lambda}\Big)}G^2(a,\lambda)\Bigg[\bigg(\frac{a}{2}-1\bigg)\big(G(a,\lambda)\xi\big)^{\frac{a}{2}-2}-\lambda\big(G(a,\lambda)\xi\big)^{\frac{a}{2}+\lambda-2}\Bigg]e^{-\big(G(a,\lambda)\xi\big)^\lambda} \\
&= \Bigg[\bigg(\frac{a}{2}-1\bigg)-\lambda\big(G(a,\lambda)\xi\big)^{\lambda}\Bigg]\frac{\phi(\xi)}{\xi}
\end{aligned}
\end{equation*}
for $G(a,\lambda) =\Gamma\big(\frac{a+2}{2\lambda}\big)/\Gamma\big(\frac{a}{2\lambda}\big)$. Substituting in $b(x/y)$, we have that
\begin{equation*}
\begin{aligned}
\int_\xi^{\infty}\phi(\eta)\eta^{\lambda-1}b\bigg(\frac{\xi}{\eta}\bigg)d\eta &= \frac{\lambda}{\Gamma\Big(\frac{a}{2\lambda}\Big)}G(a,\lambda)\bigg(\frac{a}{2}+1\bigg)\big(G(a,\lambda)\xi\big)^{\frac{a}{2}-1}\int_\xi^{\infty}\eta^{\lambda-1}e^{-\big(G(a,\lambda)\eta\big)^\lambda}d\eta \\
&= \frac{1}{\Gamma\Big(\frac{a}{2\lambda}\Big)}\Big(G(a,\lambda)\Big)^{1-\lambda}\bigg(\frac{a}{2}+1\bigg)\big(G(a,\lambda)\xi\big)^{\frac{a}{2}-1}e^{-\big(G(a,\lambda)\xi\big)^\lambda} \\
&= \frac{1}{\lambda}\Big(G(a,\lambda)\Big)^{-\lambda}\bigg(\frac{a}{2}+1\bigg)\phi(\xi)
\end{aligned}
\end{equation*}
Supplementary Equation~\eqref{sep2} may be solved to obtain
\begin{equation*}
\begin{aligned}
\omega\Bigg(2\phi(\xi)+\Bigg[\bigg(\frac{a}{2}-1\bigg)-\lambda\big(G(a,\lambda)\xi\big)^{\lambda}\Bigg]\phi(\xi)\Bigg) &= -\xi^{\lambda}\phi(\xi) + \frac{1}{\lambda}\Big(G(a,\lambda)\Big)^{-\lambda}\bigg(\frac{a}{2}+1\bigg)\phi(\xi), \\
\omega\Bigg(\frac{a}{2} + 1 - \lambda\big(G(a,\lambda)\xi\big)^{\lambda}\Bigg)\phi(\xi) &= \frac{1}{\lambda}\Big(G(a,\lambda)\Big)^{-\lambda}\Bigg(\frac{a}{2} + 1 - \lambda\big(G(a,\lambda)\xi\big)^{\lambda}\Bigg)\phi(\xi)
\end{aligned}
\end{equation*}
which is solved for all $\xi$ when
\begin{equation*}
\omega = \frac{1}{\lambda}\Big(G(a,\lambda)\Big)^{-\lambda}.
\end{equation*}
Moving to Supplementary Equation~\eqref{sep1}, we therefore have that
\begin{equation*}
\begin{aligned}
\dot{s}s^{-(\lambda+1)} &= -\frac{1}{\lambda}\Big(G(a,\lambda)\Big)^{-\lambda}, \\
\int_{s_0}^s s'^{-(\lambda+1)}ds' &= -\int_0^t \frac{1}{\lambda}\Big(G(a,\lambda)\Big)^{-\lambda}dt', \\
-\frac{1}{\lambda}\Big(s^{-\lambda}-s_0^{-\lambda}\Big) &= -\frac{1}{\lambda}\Big(G(a,\lambda)\Big)^{-\lambda}t, \\
s &= \Bigg(\Big(G(a,\lambda)\Big)^{-\lambda}t+s_0^{-\lambda}\Bigg)^{-1/\lambda}
\end{aligned}
\end{equation*}
noting that $\lambda>0$ in the case considered. The initial condition $s_0=1$ may be substituted for a fragmentation process originating from a single facet. However, in the limit of large $t$, the dependence of the typical area becomes insensitive to the initial condition, and our result may be simplified to $s(t) = \Big(G(a,\lambda)\Big)t^{-1/\lambda}$.

For the special case of $\lambda=1/2$ considered in the main results of this work,
\begin{subequations}
\begin{align}
G\Bigg(a,\lambda=\frac{1}{2}\Bigg) &= a(a+1), \label{G_supp}\\
\phi(\xi) &= \frac{a(a+1)}{2\Gamma(a)}\big(a(a+1)\xi\big)^{\frac{a}{2}-1}e^{-\sqrt{a(a+1)\xi}}, \label{phi_supp} \\
s(t) &= \frac{a(a+1)}{t^2}. \label{mean_supp}
\end{align}
\end{subequations}

\section*{\label{sec:Displacement of a 1-D random walk}Supplementary Note 2\protect}
\noindent\textbf{Displacement of a 1-D random walk.} For a random walk in one dimension comprised of random displacements $R_i$, the displacement from the origin after $2k$ steps is
\begin{equation*}
D_{2k} = \sum_{i=1}^{2k} R_i
\end{equation*}
If consecutive steps occur in opposite directions, representing a fold, then individual steps can be grouped into $k$ right (positive) and $k$ left (negative) steps:
\begin{equation*}
D_{2k} = R_k^+ - R_k^- \quad \text{where} \quad R_k^+ = \sum_{i=1,3,5...}^{2k} R_i,\quad R_k^- = \sum_{i=2,4,6...}^{2k} |R_i|
\end{equation*}
for each $r_i$ drawn from the same distribution.
If segment lengths $|R_i|$ are drawn from a gamma distribution with shape parameter $a+1$ and scale parameter $\theta$, consistent with the distribution of facet lengths traversed by a one-dimensional vertical cross-section, then $R_k^+$ and $R_k^-$ are each distributed according to a gamma distribution with shape parameter $k(a+1)$ and scale parameter $\theta$:
\begin{equation*}
|R_i| \sim \Gamma\big(a+1,\theta\big),\quad R_k^+ \sim \Gamma\big(k(a+1),\theta\big),\quad R_k^- \sim \Gamma\big(k(a+1),\theta\big)
\end{equation*}
Thus $D_{2k}$ is the difference of two identically distributed gamma variates. We can obtain the probability density function for $D_{2k}$ through a convolution of the probability density functions of $R_k^+$ and $-R_k^-$. Let $X=R_k^+$, $Y=R_k^-$, and $Z=D_{2k}$; then
\begin{equation*}
\begin{aligned}
f_Z(z) = f_{X-Y}(z) &= \int_{-\infty}^{\infty}f_X(x)f_{(-Y)}(z-x)dx \\
&= \int_{-\infty}^{\infty}f_{X}(x)f_{Y}(x-z)dx.
\end{aligned}
\end{equation*}
As $f_X(x)$ and $f_Y(y)$ both have non-negative support,
\begin{equation*}
f_Z(z) = \begin{dcases}
  \int_{0}^{\infty}f_{X}(x)f_{Y}(x-z)dx & \qquad \text{for $z \leq 0$,} \\
  \int_{0}^{\infty}f_{X}(y+z)f_{Y}(y)dy & \qquad \text{for $z > 0$,}
   \end{dcases}
\end{equation*}
where we have chosen the integration variable in the convolution to ensure the arguments of the probability density functions remain positive. With identical gamma distributions
\begin{equation*}
\begin{aligned}
f_X(x) &= \frac{1}{\theta\Gamma(k(a+1))}\bigg(\frac{x}{\theta}\bigg)^{k(a+1)-1}e^{-x/\theta}, \\
f_Y(y) &= \frac{1}{\theta\Gamma(k(a+1))}\bigg(\frac{y}{\theta}\bigg)^{k(a+1)-1}e^{-y/\theta},
\end{aligned}
\end{equation*}
where $\Gamma(k)$ is the gamma function,
\begin{equation*}
f_Z(z) = \begin{dcases}
  \frac{e^{z/\theta}}{\theta^2\Gamma(k(a+1))^2}\int_{0}^{\infty}\bigg(\frac{x}{\theta}\bigg)^{k(a+1)-1}\bigg(\frac{x-z}{\theta}\bigg)^{k(a+1)-1}e^{-2x/\theta}dx & \qquad \text{for $z \leq 0$,}\\
  \frac{e^{-z/\theta}}{\theta^2\Gamma(k(a+1))^2}\int_{0}^{\infty}\bigg(\frac{y}{\theta}\bigg)^{k(a+1)-1}\bigg(\frac{y+z}{\theta}\bigg)^{k(a+1)-1}e^{-2y/\theta}dy & \qquad \text{for $z > 0$.}
   \end{dcases}
\end{equation*}
The integral above may be solved using the following identity~\cite{gradshteyn1988tables}:
\begin{equation*}
\int_0^\infty x^{\nu-1}(x+\beta)^{\nu-1}e^{-\mu{x}}dx = \frac{1}{\sqrt{\pi}}\bigg(\frac{\beta}{\mu}\bigg)^{\nu-\frac{1}{2}}e^{\beta\mu/2}\Gamma(\nu)K_{\frac{1}{2}-\nu}\bigg(\frac{\beta\mu}{2}\bigg),
\end{equation*}
where $K_{\nu}(z)$ is the modified Bessel function of the second kind of order $\nu$. This gives
\begin{equation} \label{displacement_exact}
f_Z(z) = \frac{1}{\sqrt{\pi}\theta\Gamma(k(a+1))}\bigg(\frac{|z|}{2\theta}\bigg)^{k(a+1)-\frac{1}{2}}K_{\frac{1}{2}-k(a+1)}\bigg(\frac{|z|}{\theta}\bigg).
\end{equation}
$f_Z(z)$ should be a valid probability density function, and we can verify it indeed integrates to $1$ over its support $z\in [0,\infty)$ using the following identity~\cite{besselint2}:
\begin{equation*}
\int_0^{\infty}t^{\alpha-1}K_\nu(t)dt = 2^{\alpha-2}\Gamma\bigg(\frac{\alpha-\nu}{2}\bigg)\Gamma\bigg(\frac{\alpha+\nu}{2}\bigg).
\end{equation*}
By symmetry about $z=0$ we can integrate the following:
\begin{equation*}
\begin{aligned}
\frac{2}{\sqrt{\pi}\Gamma(k(a+1))}\bigg(\frac{1}{2}\bigg)^{k(a+1)-\frac{1}{2}}&\int_0^{\infty}\bigg(\frac{z}{\theta}\bigg)^{k(a+1)-\frac{1}{2}}K_{\frac{1}{2}-k(a+1)}\bigg(\frac{z}{\theta}\bigg)d\bigg(\frac{z}{\theta}\bigg) \\
&= \frac{2}{\sqrt{\pi}\Gamma(k(a+1))}\bigg(\frac{1}{2}\bigg)^{k(a+1)-\frac{1}{2}}2^{k(a+1)-\frac{3}{2}}\Gamma\bigg(\frac{2k(a+1)}{2}\bigg)\Gamma\bigg(\frac{1}{2}\bigg) \\
&= \frac{\Gamma(k(a+1))\Gamma(1/2)}{\sqrt{\pi}\Gamma(k(a+1))} = 1
\end{aligned}
\end{equation*}
as $\Gamma(1/2) = \sqrt{\pi}$. Furthermore, for gamma-distributed steps, the average segment length is given by $(a+1)\theta$. Thus, in our system of a one-dimensional folded strip of total length $L_0$, $k$ and $\theta$ are related as
\begin{equation*}
k = \frac{L_0}{2(a+1)\theta}
\end{equation*}
for a strip folded into $2k$ segments. Note that the total length of the walk is distributed as
\begin{equation*}
\sum_{i=1}^{2k}|R_i| \sim \Gamma(2k(a+1),\theta)
\end{equation*}
and thus has mean $2k(a+1)\theta=L_0$ and variance $2k(a+1)\theta^2=L_0\theta$ which tends to zero for small step sizes, improving the approximation of total length.

By using the following identity~\cite{besselint},
\begin{equation*}
\begin{aligned}
\int z^{-\nu}K_\nu(z)dz
&= -2^{-\nu-1}\pi z\csc(\pi\nu)\bigg[\frac{4^\nu z^{-2\nu}}{(2\nu-1)\Gamma(1-\nu)}{}_1F_2\bigg(\frac{1}{2}-\nu;1-\nu;\frac{3}{2}-\nu;\frac{z^2}{4}\bigg) \\
&+ \frac{1}{\Gamma(\nu+1)}{}_1F_2\bigg(\frac{1}{2};\frac{3}{2};\nu+1;\frac{z^2}{4}\bigg)\bigg] + \text{constant},
\end{aligned}
\end{equation*}
$f_Z(z)$ may be integrated analytically to obtain an expression for a two-sided survival function as
\begin{equation} \label{survival_fun_exact}
\begin{aligned}
S_Z(w;a,\theta) &= P(|Z|>w;w\geq0) \\
&= 1-2\int_0^{w}f_Z(z)dz \\
&=1+\frac{\sqrt{\pi}}{\Gamma(k(a+1))}\bigg(\frac{w}{\theta}\bigg)\csc(\pi\nu)\Bigg[\frac{4^\nu}{(2\nu-1)\Gamma(1-\nu)}\bigg(\frac{w}{\theta}\bigg)^{-2\nu}{}_1F_2\bigg(\frac{1}{2}-\nu;1-\nu,\frac{3}{2}-\nu;\frac{1}{4}\Big(\frac{w}{\theta}\Big)^2\bigg)\\
&+\frac{1}{\Gamma(1+\nu)}{}_1F_2\bigg(\frac{1}{2};\frac{3}{2},1+\nu;\frac{1}{4}\Big(\frac{w}{\theta}\Big)^2\bigg)\Bigg],
\end{aligned}
\end{equation}
where $\nu=\frac{1}{2}-k(a+1)$.
\section*{\label{sec:Asymptotic Approximation}Supplementary Note 3\protect}
\noindent\textbf{Asymptotic Approximation.} The main text considers the limit of large $k$, when the number of steps is large and the step size is small, which permits application of the central limit theorem such that
\begin{equation*}
    R_k^+, R_k^- \sim \mathcal{N}\Big(k(a+1)\theta,k(a+1)\theta^2\Big),
\end{equation*}
where $\mathcal{N}(\mu,\sigma^2)$ is a normal distribution with mean $\mu$ and variance $\sigma^2$.
Then, the displacement from the origin is
\begin{equation*}
    D_{2k} = R_k^+ - R_k^- \sim \mathcal{N}(0,2k(a+1)\theta^2) = \mathcal{N}(0,L_0\theta).
\end{equation*}
Thus, in the limit of large $k$,
\begin{equation} \label{displacement_approx}
    f_Z(z;\theta) = \frac{1}{\sqrt{2\pi{L_0}\theta}}\exp\Bigg(-\frac{z^2}{2L_0\theta}\Bigg),
\end{equation}
and the corresponding survival function, for $w\geq 0$, is
\begin{equation} \label{survival_fun_approx}
S_Z(w;\theta) = 1-\text{erf}\Bigg(\frac{w}{\sqrt{2L_0\theta}}\Bigg),
\end{equation}
where $\text{erf}(z)$ is the error function. This approximation is valid for strongly crumpled experimental samples, but agreement is not guaranteed for samples in the large $\tilde{\Delta}$ regime which have few facets. Thus, we found it important to carry out the exact derivation of Supplementary Equations~\eqref{displacement_exact} \&~\eqref{survival_fun_exact}  to ensure consistency with their respective approximations, Supplementary Equations~\eqref{displacement_approx} \&~\eqref{survival_fun_approx}. Supplementary Fig.~\ref{num_rw_approx} shows the increasing agreement of Supplementary Equation~\eqref{survival_fun_exact} and the approximation given by Supplementary Equation~\eqref{survival_fun_approx} for large $k$.
 
Next, as explained in the main text, the incremental change in $t$ with crumpling iteration behaves as
 \begin{equation}\label{prog_eqn_supp}
\delta{t} \equiv \frac{\partial{t}}{\partial{n}} =  \alpha\frac{1-\tilde{\Delta}}{\tilde{\Delta}}S_Z(w;\theta),
\end{equation}
where $\alpha$ is a constant of proportionality, and $\theta=1/t$. Once again making use of asymptotic approximations, we can integrate $\delta{t}/S_Z(w;t)$ as follows:
In the limit of large $t$,
\begin{equation*}
    S_Z(w;t) = 1-\text{erf}\Bigg(\frac{w}{\sqrt{2L_0/t}}\Bigg) \approx \frac{e^{-w^2t/2L_0}}{w\sqrt{\pi t/2L_0}}.
\end{equation*}
By a change of variables $u=w\sqrt{t/2L_0}$,
\begin{equation*}
    I = \int_0^t \frac{dt'}{S_Z(w;t')} = \frac{4L_0\sqrt{\pi}}{w^2}\int_0^u u'^2e^{u'^2} du',
\end{equation*}
which to leading order in $u$ yields
\begin{equation*}
    \begin{aligned}
    I &\approx \frac{2L_0\sqrt{\pi}}{w^2}ue^{u^2} \\
    &= \alpha\frac{1-\tilde{\Delta}}{\tilde{\Delta}}n + c
    \end{aligned}
\end{equation*}
by consequence of Supplementary Equation~\eqref{prog_eqn_supp}, where $c$ is an integration constant.
In order to solve for $u$, we recall the definition of the Lambert $W$ function, or product logarithm, which gives the inverse solution
\begin{equation*}
    x = W_0(y)
\end{equation*}
to
\begin{equation*}
    y = xe^x,
\end{equation*}
where $W_0(y)$ is the principal branch of the Lambert $W$ function valid for real $x$ and $y$, and positive $y$. Defining a new variable $z=\Big(\alpha\frac{1-\tilde{\Delta}}{\tilde{\Delta}}n + c\Big)w^2/L_0\sqrt{2\pi}$, we obtain
\begin{equation*}
    u = \sqrt{W_0(z^2)/2}
\end{equation*}
or, expressed in terms of $t$,
\begin{equation*}
    t = \frac{L_0}{w^2}W_0(z^2).
\end{equation*}
Making a final asymptotic approximation, $W_0(y) \approx \log(y)$, we thus have that
\begin{equation*}
\begin{aligned}
    t &\approx \frac{2L_0}{w^2}\log(z) \\
    &= \frac{2L_0}{w^2}\log\Bigg(\frac{w^2}{L_0\sqrt{2\pi}}\Big(\alpha\frac{1-\tilde{\Delta}}{\tilde{\Delta}}n + c\Big)\Bigg).
\end{aligned}
\end{equation*}
With the condition $t(n=0)=0$, we obtain the final relation
\begin{equation}
    t(n,\tilde{\Delta};w) = \frac{2L_0}{w^2}\log\Bigg(1 + \frac{\alpha w^2}{L_0\sqrt{2\pi}}\frac{1-\tilde{\Delta}}{\tilde{\Delta}}n\Bigg). \label{t_vs_n}
\end{equation}
 
\section*{\label{sec:Critical confinement}Supplementary Note 4\protect}
\noindent\textbf{Critical confinement.} Our model of a folded one-dimensional strip as a random walk relates geometric incompatibility to the random walk stepping outside a confinement distance $w$. This critical distance $w$ is dictated by the geometry of the imposed confinement, and the way in which the one-dimensional strip folds into \textit{stacks} of one or more folded \textit{layers} to accommodate its full length within the allowed space. Let $m$ represent the number of spaced stacks, and $p$ the average number of layers per stack, in our strip of length $L_0$, confined to a rectangular container of width $R$ and height $L$, $L\leq L_0$. To satisfy the constraint of total length $L_0$ at any compaction $\tilde{\Delta}=L/L_0$, we must have $mp\sqrt{(L/m)^2 + R^2} = L_0$. The critical width $w$ of facets which would fragment under further confinement is given by $w = L_0/mp = \sqrt{(L/m)^2 + R^2}$.
At low confinement, $p \approx 1$, and thus our constraint gives $m=(L_0/R)\sqrt{1-\tilde{\Delta}^2}$, resulting in
\begin{equation}\label{conf}
w(\tilde{\Delta}) = \frac{R}{\sqrt{1-\tilde{\Delta}^2}}.
\end{equation}
At high confinement, the collapse of stacks leads to a decrease in $m$ that scales in proportion to $L$, in turn scaling the number of layers $p \sim 1/L$. Specifically, we can define
\begin{equation*}
\begin{aligned}
m(\tilde\Delta\rightarrow 0) \sim \frac{L}{R}, \\
p(\tilde\Delta\rightarrow 0) \sim \frac{L_0}{L}
\end{aligned}
\end{equation*}
and obtain
\begin{equation*}
w(\tilde\Delta\rightarrow 0) = \frac{L_o}{mp} \sim R
\end{equation*}
which is consistent with our result at low confinement taken to the limit of small $L$. Thus we use Supplementary Equation~\eqref{conf} throughout.
Substituting Supplementary Equation~\eqref{conf} into Supplementary Equation~\eqref{t_vs_n}, we arrive at an expression for $t$ solely in terms of $n$ and $\tilde{\Delta}$:
\begin{equation}
    t(n,\tilde{\Delta}) = \tilde{c}_1(1-\tilde{\Delta}^2)\log\Bigg(1+\frac{\tilde{c}_2n}{\tilde{\Delta}(1+\tilde{\Delta})}\Bigg),
\end{equation}
where $\tilde{c}_1 = 2L_0/R^2$ and $\tilde{c}_2 = \alpha R^2/L_0\sqrt{2\pi}$.

\end{document}